\documentclass[useAMS,usenatbib]{mn2e}

\usepackage{amsmath, amssymb}
\usepackage[dvips]{graphicx}
\usepackage{epsfig}

\newcommand{\rot}{\bmath{\nabla} \times}
\newcommand{\divg}{\bmath{\nabla}\cdot}

\newcommand{\rlight}{r_{\rm L}}
\newcommand{\Rs}{R_{\rm s}}

\newcommand{\ez}{\bmath{e}_{\rm z}}
\newcommand{\er}{\bmath{e}_{\rm r}}
\newcommand{\etheta}{\bmath{e}_\vartheta}
\newcommand{\ephi}{\bmath{e}_\varphi}

\newcommand{\aap}{A\&A}
\newcommand{\mnras}{MNRAS}

\newcommand{\apj}{ApJ}
\newcommand{\apjl}{ApJL}

\newcommand{\prd}{Physical Review D}

\title[Monopole fields around neutron stars]{General-relativistic monopole magnetosphere of neutron stars: a pseudo-spectral discontinuous Galerkin approach} 
\author[J. P\'etri]{J.  P\'etri$^{1}$
\thanks{E-mail: jerome.petri@astro.unistra.fr} \\
  $^{1}$Observatoire Astronomique de Strasbourg, Universit\'e de Strasbourg, CNRS, UMR 7550, 11 rue de l'Universit\'e, F-67000 Strasbourg, France.}

\begin{document}

\date{Accepted . Received ; in original form }

\pagerange{\pageref{firstpage}--\pageref{lastpage}} 
\pubyear{2014}

\maketitle

\label{firstpage}

\begin{abstract}
The close vicinity of neutron stars remains poorly constrained by observations. Although plenty of data are available for the peculiar class of pulsars we are still unable to deduce the underlying plasma distribution in their magnetosphere. In the present paper, we try to unravel the magnetospheric structure starting from basic physics principles and reasonable assumptions about the magnetosphere. Beginning with the monopole force-free case, we compute accurate general-relativistic solutions for the electromagnetic field around a slowly rotating magnetized neutron star. Moreover, here we address this problem by including the important effect of plasma screening. This is achieved by solving the time-dependent Maxwell equations in a curved space-time following the 3+1~formalism. We improved our previous numerical code based on pseudo-spectral methods in order to allow for possible discontinuities in the solution. Our algorithm based on a multi-domain decomposition of the simulation box belongs to the discontinuous Galerkin finite element methods. We performed several sets of simulations to look for the general-relativistic force-free monopole and split monopole solutions. Results show that our code is extremely powerful in handling extended domains of hundredth of light-cylinder radii~$\rlight$. The code has been validated against known exact analytical monopole solutions in flat space-time. We also present semi-analytical calculations for the general-relativistic vacuum monopole.
\end{abstract}


\section{INTRODUCTION}

It is well admitted that pulsars are strongly magnetized and rotating neutron stars surrounded by electron-positron pairs filling their magnetosphere. However an accurate description of the interaction between this plasma and the neutron star electromagnetic field remains poorly constrained by observations. Moreover a realistic model should also include some radiative processes. We are still far from a comprehensive and self-consistent picture of the pulsar machinery. Both plasma flows and strong gravity impact on the structure of the magnetosphere. Curvature and frame-dragging effects are indeed important due to the high compactness of neutron stars. For typical models of neutron star interiors, the compactness is about $\Xi = \Rs/R \approx 0.5$ where $\Rs=2\,G\,M/c^2$ is the Schwarzschild radius, $M$ is the mass of the neutron star, $R$ its radius, $G$ the gravitational constant and $c$ the speed of light.

It is the purpose of this paper to study the response of the electromagnetic field to the combined effect of plasma screening and curved space-time in the vicinity of a neutron star. To this aim we compute general-relativistic solutions in the force-free approximation. The problem we therefore address is similar to the electrodynamics of black hole magnetospheres. Actually, the numerical technique employed are the same expected that in our case we do not have any complication arising from the presence of an event horizon. \cite{2004MNRAS.350.1431K} was the first to report on numerical simulations of general-relativistic monopole magnetospheres of black holes in the magnetohydrodynamic regime. \cite{2004MNRAS.350..427K} also investigated the properties of the magnetospheric plasma in the force-free limit. Since then, several authors followed the effort of modelling general-relativistic magnetospheres of compact objects. \cite{2006MNRAS.367.1797M} designed a general-relativistic code for force-free magnetospheres and \cite{2006MNRAS.368L..30M} applied it also to neutron stars.

\cite{2013MNRAS.433..986P} showed that multipole vacuum solutions in general relativity can be computed semi-analytically via the 3+1~formalism through a vector spherical harmonic expansion method introduced by \cite{2012MNRAS.424..605P}. However it is well known that a neutron star cannot be surrounded by vacuum. Indeed, the electric field induced by the rotation of the magnetic field generates huge Lorentz forces able to extract particles from the crust and therefore filling the magnetosphere. For simplicity, as a first step towards more realistic magnetospheres, the force-free assumption is often quoted. In that case the plasma dynamics is completely dominated by the electromagnetic field which is a good approximation for neutron star magnetospheres. The resulting force-free geometry has been investigated by many authors like for instance in the aligned case by \cite{1999ApJ...511..351C, 2012MNRAS.423.1416P} and the general oblique rotator by \cite{2006ApJ...648L..51S, 2009A&A...496..495K, 2012MNRAS.420.2793K, 2012MNRAS.424..605P}. General-relativistic force-free neutron star magnetospheres have been less investigated so far. But \cite{1990SvAL...16..286B} already mentioned that general-relativistic effects can significantly distort the parallel component (with respect to the magnetic field) of the electric field. This can have important implications for particle creation, acceleration and radiation in the polar caps. Indeed, deviation from the corotation charge density leads to a parallel component of the electric field determined by the magnetic field geometry. Therefore, as also claimed by \cite{1992MNRAS.255...61M}, space-time curvature and frame dragging effects are important for the electrodynamics of the gaps. Several numerical techniques have been applied to model such magnetospheres. Usually the schemes are closely related to the finite volume algorithm, a well tested method for computational fluid dynamics due to its conservative properties. High resolution shock capturing techniques enable an increase of the spatial order of the method but at the expense of larger stencils. Such scheme are also useful to solve Maxwell equations. Recently, another arbitrary high order method, the discontinuous Galerkin approach, has been tested in general relativity by \cite{2011PhRvD..84b4010R}.

Our goal in this paper is to quantify precisely the distortion induced by general-relativistic effects, namely curvature of space-time and frame dragging. To this end, we solve the time-dependent Maxwell equations in curved space-time in spherical coordinates. Nevertheless, as a starting point we restrict the solutions to the monopole field in order to elucidate the consequences of general relativity avoiding complications induced for instance by the presence of a cusp at the light-cylinder in the case of an aligned dipolar magnetic field. Strictly speaking, at this Y-point, the magnetic field strength vanishes and can lead to problems in the force-free approximation due to the electric current density prescription. Nevertheless, in order to show how efficiently the code can handle discontinuities such as current sheets for instance in the equatorial plane, we present Newtonian as well as general-relativistic simulations of the split monopole field. Consequently, we use the 3+1~formalism of electrodynamics as briefly reminded in Section~\ref{sec:Modele}. Next we give approximate solutions to the vacuum monopole field in Section~\ref{sec:Monopole} which will be useful for benchmarking the code whose algorithm is described in Section~\ref{sec:Algorithme} and then tested in flat space-time in Section~\ref{sec:Tests}. Application of our new code to vacuum and force-free curved space-time monopoles are presented in Section~\ref{sec:Results}. We extend our study to the split monopole case to demonstrate the ease of handling discontinuities. Conclusions and ongoing work are drawn in Section~\ref{sec:Conclusion}.

\section{The 3+1~formalism}
\label{sec:Modele}

In this section we briefly remind the set of Maxwell equations in curved space-time following the 3+1~formalism for a fixed background metric. We split space-time into an absolute space~$\{x,y,z\}$ and a universal time~$t$, similar to our all day experience. Advantages of such a split have been demonstrated in many numerical simulations about neutron stars and black hole magnetospheres.

\subsection{The split of the space-time metric}

The four dimensional space-time is split into a 3+1~foliation such that the background metric~$g_{ik}$ can be expressed as
\begin{equation}
  \label{eq:metrique}
  ds^2 = g_{ik} \, dx^i \, dx^k = \alpha^2 \, c^2 \, dt^2 - \gamma_{ab} \, ( dx^a + \beta^a \, c\,dt ) \, (dx^b + \beta^b \, c\,dt )
\end{equation}
where $x^i = (c\,t,x^a)$, $t$ is the time coordinate or universal time and $x^a$ some associated space coordinates. The Landau-Lifschitz convention is used for the metric signature given by $(+,-,-,-)$ \citep{LandauLifchitzTome2}. $\alpha$ is the lapse function, $\beta^a$ the shift vector and $\gamma_{ab}$ the spatial metric of absolute space.  By convention, latin letters from $a$ to $h$ are used for the components of vectors in absolute space, in the range~$\{1,2,3\}$, whereas latin letters starting from $i$ are used for four dimensional vectors and tensors, in the range~$\{0,1,2,3\}$. A fiducial observer (FIDO) is defined by its 4-velocity~$n^i$ such that
\begin{subequations}
 \begin{align}
  n^i & = \frac{dx^i}{d\tau} = \frac{c}{\alpha} \, ( 1, - \bbeta) \\
  n_i & = (\alpha \, c, \bmath{0}) 
 \end{align}
\end{subequations}
This vector is orthogonal to the hyper-surface of constant time coordinate~$\Sigma_t$. Its proper time~$\tau$ is measured according to
\begin{equation}
  d\tau = \alpha\,dt
\end{equation}
For a slowly rotating neutron star, the lapse function is
\begin{equation}
  \label{eq:Lapse}
  \alpha = \sqrt{ 1 - \frac{\Rs}{r} }
\end{equation}
and the shift vector
\begin{subequations}
 \begin{align}
  \label{eq:Shift}
  c \, \bbeta = & - \omega \, r \, \sin\vartheta \, \ephi \\
  \omega = & \frac{\Rs\,a\,c}{r^3}
 \end{align}
\end{subequations}
We use a spherical coordinate system~$(r,\vartheta,\varphi)$ and an orthonormal spatial basis~$(\er,\etheta,\ephi)$. The metric of a slowly rotating neutron star remains close to the usual flat space, except for the radial direction. Indeed the components of the spatial metric are given in Boyer-Lindquist coordinates by
\begin{equation}
  \label{eq:Metric3D}
  \gamma_{ab} =
  \begin{pmatrix}
    \alpha^{-2} & 0 & 0 \\
    0 & r^2 & 0 \\
    0 & 0 & r^2 \sin^2\vartheta
  \end{pmatrix}
\end{equation}
For this slow rotation approximation, the spatial metric does not depend on the spin frequency of the massive body but only on $M$ through $\alpha$. The spin~$a$ is related to the angular momentum~$J$ by $J=M\,a\,c$. It follows that $a$ has units of a length and should satisfy $a \leq \Rs/2$. Introducing the moment of inertia~$I$, we also have $J=I\,\Omega$ , $\Omega$ being the spin frequency. In the special case of a homogeneous and uniform neutron star interior with spherical symmetry, the moment of inertia reads
\begin{equation}
  \label{eq:Inertie}
  I = \frac{2}{5} \, M \, R^2
\end{equation}
Thus the spin parameter can be expressed as
\begin{equation}
  \label{eq:spin}
  \frac{a}{\Rs} = \frac{2}{5} \, \frac{R}{\Rs} \, \frac{R}{\rlight}
\end{equation}
For the remainder of this paper, we will use this expression to relate the spin parameter intervening in the metric to the spin frequency of the neutron star. From the above expression, note that the parameter~$a/\Rs$ remains smaller than~0.4 because $R=2\,\Rs$ and $\rlight\geq2\,R$ in our set of simulations.

\subsection{Maxwell equations}

Maxwell equations in absolute space take a form very similar to their traditional expression in Newtonian space except that space is curved. The time-dependent Maxwell equations in a prescribed metric (possibly time-dependent) read
\begin{subequations}
\begin{align}
\label{eq:Maxwell1}
 \divg \bmath B & = 0 \\
\label{eq:Maxwell2}
 \rot \bmath E & = - \frac{1}{\sqrt{\gamma}} \, \partial_t (\sqrt{\gamma} \, \bmath B) \\
\label{eq:Maxwell3}
 \divg \bmath D & = \rho \\
\label{eq:Maxwell4}
 \rot \bmath H & = \bmath J + \frac{1}{\sqrt{\gamma}} \, \partial_t (\sqrt{\gamma} \, \bmath D)
\end{align}
\end{subequations}
The source terms $(\rho, \bmath J)$ will be specified by the force-free condition, see paragraph below. The three dimensional vector fields are not independent, they are related by two important constitutive relations, namely
\begin{subequations}
\begin{align}
\label{eq:ConstitutiveE}
  \varepsilon_0 \, \bmath{E} & = \alpha \, \bmath{D} + \varepsilon_0\,c\,\bbeta \times \bmath{B} \\
\label{eq:ConstitutiveH}
  \mu_0 \, \bmath{H} & = \alpha \, \bmath{B} - \frac{\bbeta \times \bmath{D}}{\varepsilon_0\,c}
\end{align}
\end{subequations}
$\varepsilon_0$ is the vacuum permittivity and $\mu_0$ the vacuum permeability. 
The curvature of absolute space is taken into account by the lapse function factor~$\alpha$ in the first term on the right-hand side and the frame dragging effect is included in the second term, the cross-product between the shift vector~$\bbeta$ and the fields. The derivation of the above equations is given in \cite{2004MNRAS.350..427K}. From the auxiliary fields $(\textbf{\textit{E}}, \textbf{\textit{H}})$ we get the Poynting flux through a sphere of radius $r$ by computing the two dimensional integral on this sphere by
\begin{equation}
\label{eq:Poynting}
 L = \int_\Omega \bmath{E} \wedge \bmath{H} \, r^2 \, d\Omega
\end{equation}
where $d\Omega$ is the infinitesimal solid angle and $\Omega$ the full sky angle of~$4\,\upi$~sr.

\subsection{Force-free conditions}

The source terms have not yet been specified. They are deduced from the force-free condition that in the 3+1~formalism become
\begin{subequations}
\begin{align}
  \bmath J \cdot \bmath E & = 0 \\
  \rho \, \bmath E + \bmath J \times \bmath B & = \bmath 0
\end{align}
\end{subequations}
which implies $\bmath E \cdot \bmath B = 0 $ and therefore also $\bmath D \cdot \bmath B = 0 $. As in the special relativistic case, the current density is found to be, see the derivation for instance in \cite{2011MNRAS.418L..94K}
\begin{equation}
 \label{eq:CourantFFE}
 \bmath J = \rho \, \frac{\bmath E \times \bmath B}{B^2} + \frac{\bmath B \cdot \rot \bmath H - \bmath D \cdot \rot \bmath E}{B^2} \, \bmath B
\end{equation}
$\bmath B$ and $\bmath D/\varepsilon_0$ can be interpreted as the magnetic and electric field respectively as measured by the FIDO. Moreover its electric current density~$\bmath j$ is given by
\begin{equation}
  \label{eq:CourantFIDO}
  \alpha \, \bmath j = \bmath J + \rho \, c \, \bbeta
\end{equation}
Maxwell equations~(\ref{eq:Maxwell1})-(\ref{eq:Maxwell4}), the constitutive relations in equations~(\ref{eq:ConstitutiveE}) and (\ref{eq:ConstitutiveH}) and the prescription for the source terms in equation~(\ref{eq:CourantFFE}) set the background system to be solved for any prescribed metric in the force-free approximation.

\onecolumn

\section{VACUUM MONOPOLE FIELD}
\label{sec:Monopole}

Before dealing with the force-free solution, we recall the exact vacuum electromagnetic field in flat space-time and extend the result to the general-relativistic monopole field, valid up to first order in the spin parameter of the star. Although the monopole assumption is not realistic for the zeroth order magnetic field of the neutron star, it gives us insight into the effects of curved space-time on to force-free magnetospheres. Such solutions will also serve as benchmark for testing and checking current and forthcoming electromagnetic codes in general relativity.

\subsection{Newtonian solution}

We start with a simple monopole magnetic field anchored in a perfectly conducting star of radius~$R$ and rotating at a speed~$\Omega$ around an axis passing through its centre. Let us denote this axis by $\ez$. The strength of the magnetic field at the surface is $B$. Thus, in Minkowski space-time, the exterior vacuum solution for a rotating magnetic monopole is given by 
\begin{subequations}
 \begin{align}
 \label{eq:MonopoleB}
  \bmath B & = B \, \frac{R^2}{r^2} \, \er \\
 \label{eq:MonopoleE}
  \bmath E & = - \frac{\Omega\,B\,R^4}{r^3} \, ( 2 \, \cos\vartheta \, \er + \sin\vartheta \, \etheta )
 \end{align}
\end{subequations}
assuming that the electric field in the comoving frame vanishes in the interior of the star. The induced electric field is therefore of dipolar nature. Note that the relation between $\bmath E$ and $\bmath D$ is simply $\varepsilon_0 \, \bmath E = \bmath D$. In terms of the "potential", see equation~(\ref{eq:potentielD}) below, we can write it as
\begin{equation}
 \bmath E = 2 \, \sqrt{\frac{2\,\upi}{3}} \, \Omega\,B\,R^4 \, \mathrm{Re} \left[ \rot \left( \frac{\bmath \Psi_{1,0}}{r^2} \right) \right]
\end{equation}
This means that the only non-vanishing coefficient is
\begin{equation}
\label{eq:f10E}
 f^E_{1,0} = 2 \, \sqrt{\frac{2\,\upi}{3}} \, \frac{\Omega\,B\,R^4}{r^2}.
\end{equation}
All other coefficients of the expansion like $f^E_{l,m}$ and $g^E_{l,m}$ should be equal to zero. Remember that a divergencelessness vector field $\bmath E$ can be expanded according to
\begin{equation}
  \label{eq:Decomposition_HSV_div_0_E}
  \mathbf{E}(r,\vartheta,\varphi,t) = \sum_{l=1}^\infty\sum_{m=-l}^l \left( \rot [f^E_{l,m}(r,t) \, \mathbf{\Phi}_{l,m}] + g^E_{l,m}(r,t) \, \mathbf{\Phi}_{l,m} \right) \\
\end{equation}
where $\mathbf{\Phi}_{l,m}$ are vector spherical harmonics, see for instance \cite{2013MNRAS.433..986P}.
Later, we will use this expression to check the numerical accuracy of our code, see section~\ref{sec:Tests}.

\subsection{General-relativistic solution}

In order to look for the analytical solution to the general-relativistic monopole field in vacuum, we use the formalism developed in depth by \cite{2013MNRAS.433..986P}. Closed analytical expressions have only been found for the first order expansion of the electric field~$\bmath D$ as described in the first part of this section. For higher order approximations, we have to resort to numerical solutions which are exposed in the second part of this section.

\subsubsection{First order expansion}

The background monopolar magnetic field in eq.~(\ref{eq:MonopoleB}) remains exact for the curved space-time geometry. We are looking for a first order approximation to the electric field such that
\begin{equation}
\label{eq:potentielD}
   \bmath D = \mathrm{Re} \left[ \rot ( f_{1,0}^D \, \bmath \Phi_{1,0} ) \right]
\end{equation}
which automatically satisfies $\divg \bmath{D}=0$. To zeroth order, the magnetic field is not perturbed, we leave it unchanged. For the electric field, the function~$f_{1,0}^D$ satisfies
\begin{equation}
 \partial_r(\alpha^2\,\partial_r(r\,f_{1,0}^D)) - \frac{2}{r} \, f_{1,0}^D = -6\,\sqrt{\frac{2\,\upi}{3}} \, \varepsilon_0 \, c \, \frac{a \, B \, \Rs \, R^2}{r^4} = -6\,\sqrt{\frac{2\,\upi}{3}} \, \varepsilon_0 \, \frac{\omega \, B \, R^2}{r}
\end{equation}
A particular solution vanishing at infinity is
\begin{equation}
  \label{eq:fD10part}
  f_{1,0}^{D(p)} = 6\, \sqrt{\frac{2\,\upi}{3}} \, \varepsilon_0 \, c \, \frac{a \, B \, R^2}{\Rs^2} \, \left[ \frac{r}{\Rs} \, \ln\alpha^2 + 1 + \frac{R_s}{2\,r} + \frac{R_s^2}{3\,r^2} \right]
\end{equation}
The general solution of the homogeneous equation also vanishing at infinity is
\begin{equation}
  \label{eq:fD10hom}
  f_{1,0}^{D(h)} = K \, r \, \left[ \ln\alpha^2 + \frac{R_s}{r} + \frac{R_s^2}{2\,r^2} \right]
\end{equation}
This expression was first obtained by \cite{1964SPhD....9..329G}. The boundary condition at the stellar crust is
\begin{equation}
 \left. \frac{1}{r} \, \partial_r (r\,f_{1,0}^{D})\right|_R = -2 \, \sqrt{\frac{2\,\upi}{3}} \, \varepsilon_0 \, \frac{\tilde{\omega}_R\,B\,R}{\alpha_R^2}
\end{equation}
For the general and particular solutions we have respectively
\begin{subequations}
  \begin{align}
 \label{eq:drfD10dr}
  \frac{1}{r} \, \partial_r (r\,f_{1,0}^{D(h)}) & = K \, \left[ 2\, \ln\alpha^2 + \frac{R_s}{r} \, \frac{2\,r-\Rs}{r-\Rs} \right] \\
  \frac{1}{r} \, \partial_r (r\,f_{1,0}^{D(p)}) & = 6\, \sqrt{\frac{2\,\upi}{3}} \, \varepsilon_0 \, c \, \frac{a \, B \, R^2}{r\,\Rs^2} \, \left[ 2\,\frac{r}{\Rs} \, \ln\alpha^2 + 1 + \frac{1}{\alpha^2} - \frac{R_s^2}{3\,r^2} \right]
 \end{align}
\end{subequations}
For convenience, we introduce the following constants
\begin{subequations}
 \begin{align}
  C_1 & = 2\,\frac{R}{\Rs} \, \ln\alpha_R^2 + 1 + \frac{1}{\alpha_R^2} - \frac{R_s^2}{3\,R^2} \\
  C_2 & = \left[ 2\, \ln\alpha_R^2 + \frac{R_s}{R} \, \frac{2\,R-\Rs}{R-\Rs} \right]^{-1} \\
  \tilde{\omega}_R & = \Omega - \omega_R
 \end{align}
\end{subequations}
The index $R$ means that quantities are evaluated on the neutron star surface.
Then the constant of integration in eq.~(\ref{eq:fD10hom}) reads
\begin{equation}
 K = -2 \, \sqrt{\frac{2\,\upi}{3}} \, \varepsilon_0 \, C_2 \, B \, R \, \left[ \frac{\tilde{\omega}_R}{\alpha_R^2} + 3 \, C_1 \, \frac{\omega_R \, R^3}{\Rs^3} \right]
\end{equation}
To summarize, to first order in spin parameter, the electric field satisfies
\begin{equation}
 \label{eq:fD10}
f_{1,0}^D = K \, r \, \left[ \ln\alpha^2 + \frac{R_s}{r} + \frac{R_s^2}{2\,r^2} \right] + 6\, \sqrt{\frac{2\,\upi}{3}} \, \varepsilon_0 \, c \, \frac{a \, B \, R^2}{\Rs^2} \, \left[ \frac{r}{\Rs} \, \ln\alpha^2 + 1 + \frac{R_s}{2\,r} + \frac{R_s^2}{3\,r^2} \right]
\end{equation}
We will use this analytical expressions to check our code in the general-relativistic case to the lowest order in the spin parameter expansion.

In the limit of a weak gravitational field, the solution reduces to equation~(\ref{eq:f10E}) as it should. To the next leading order in~$a$, we expect a dipolar perturbation of the magnetic field, thus we write
\begin{equation}
 \bmath B = B \, \frac{R^2}{r^2} \, \er + \mathrm{Re} \left[ \rot ( f_{2,0}^B \, \bmath \Phi_{2,0} ) \right]
\end{equation}
The function $f_{2,0}^B$ will be a solution of
\begin{equation}
 \partial_r(\alpha^2\,\partial_r(r\,f_{2,0}^{B})) - \frac{6}{r} \, f_{2,0}^{B} = - \frac{6}{\sqrt{5}} \, \mu_0 \, \omega \, f_{1,0}^{D}
\end{equation}
Taking into account the boundary conditions, $f_{2,0}^{B}$ has to vanish at infinity and at the neutron star surface. This corresponds to the Deutsch approach where the radiative disturbances of the normal component of $\bmath B$ are not taken into account. So far we have not found any analytical expression to solve this boundary value problem. We have to resort to numerical integration. This is explained in the next paragraph.

\subsubsection{Multipole expansion}

The most general situation including multipoles to any order is exposed in this paragraph. We look for solutions that can be expanded in the following series
\begin{subequations}
\begin{align}
   \bmath D & = \mathrm{Re} \left[ \rot ( \sum_{l\geq1} f_{l,0}^D \, \bmath \Phi_{l,0} ) \right] \\
   \bmath B & = B \, \frac{R^2}{r^2} \, \er + \mathrm{Re} \left[ \rot ( \sum_{l\geq1} f_{l,0}^B \, \bmath \Phi_{l,0} ) \right]
\end{align}
\end{subequations}
Each of the coefficient $f_{l,0}^{D}$ and $f_{l,0}^{B}$ has to satisfy the differential equation which is given for the electric and magnetic field respectively by
\begin{subequations}
\begin{align}
\label{eq:fl0D}
 \partial_r(\alpha^2\,\partial_r(r\,f_{l,0}^{D})) - \frac{l(l+1)}{r} \, f_{l,0}^{D} & = 
 3 \, \varepsilon_0 \, \omega \, \left[ \sqrt{(l-1)\,(l+1)} \, J_{l,0} \, f_{l-1,0}^{B} - \sqrt{l\,(l+2)} \, J_{l+1,0} \, f_{l+1,0}^{B} \right] \nonumber \\
& - 6\,\sqrt{\frac{2\,\upi}{3}} \, \varepsilon_0 \, c \, \frac{a \, B \, \Rs \, R^2}{r^4} \, \delta_{l,1} \\
 \label{eq:fl0B}
 \partial_r(\alpha^2\,\partial_r(r\,f_{l,0}^{B})) - \frac{l(l+1)}{r} \, f_{l,0}^{B} & = 
 - 3 \, \mu_0 \, \omega \, \left[ \sqrt{(l-1)(l+1)} \, J_{l,0} \, f_{l-1,0}^{D} - \sqrt{l\,(l+2)} \, J_{l+1,0} \, f_{l+1,0}^{D} \right]
\end{align}
\end{subequations}
The Kronecker symbol~$\delta_{l,1}$ appearing in the differential equation for the electric field represents the contribution from the monopole magnetic field, that cannot be expressed in terms of a curl. We add it explicitly.

Let us write down these equations for the three first coefficients in $\bmath{B}$ and $\bmath{D}$. The system of partial differential equations then reads
\begin{subequations}
 \begin{align}
\label{eq:EDPf10D}
   \partial_r(\alpha^2\,\partial_r(r\,f_{1,0}^{D})) - \frac{2}{r} \, f_{1,0}^{D} & = - 6 \, \varepsilon_0 \, \omega \, \left[ \frac{1}{\sqrt{5}} \, f_{2,0}^{B} + \sqrt{\frac{2\,\upi}{3}} \, B \, \frac{R^2}{r} \right] \\
 \label{eq:EDPf30D}
 \partial_r(\alpha^2\,\partial_r(r\,f_{3,0}^{D})) - \frac{12}{r} \, f_{3,0}^{D} & = \frac{1}{\sqrt{7}} \, \varepsilon_0 \, \omega \, \left[ 18 \, \sqrt{\frac{2}{5}} \, f_{2,0}^{B} - 4 \, \sqrt{15} \, f_{4,0}^{B} \right] \\
 \label{eq:EDPf50D}
 \partial_r(\alpha^2\,\partial_r(r\,f_{5,0}^{D})) - \frac{30}{r} \, f_{5,0}^{D} & = \frac{2}{\sqrt{11}} \, \varepsilon_0 \, \omega \, \left[ 5 \, \sqrt{6} \, f_{4,0}^{B} - 9 \, \sqrt{\frac{35}{13}} \, f_{6,0}^{B} \right] \\
\label{eq:EDPf20B}
 \partial_r(\alpha^2\,\partial_r(r\,f_{2,0}^{B})) - \frac{6}{r} \, f_{2,0}^{B} & = - \frac{6}{\sqrt{5}} \, \mu_0 \, \omega \, \left[ f_{1,0}^{D} - 3 \sqrt{\frac{2}{7}} \, f_{3,0}^{D} \right] \\
\label{eq:EDPf40B}
 \partial_r(\alpha^2\,\partial_r(r\,f_{4,0}^{B})) - \frac{20}{r} \, f_{4,0}^{B} & = - 2 \, \sqrt{3} \, \mu_0 \, \omega \, \left[ 2 \, \sqrt{\frac{5}{7}} \, f_{3,0}^{D} - 5 \, \sqrt{\frac{2}{11}} \, f_{5,0}^{D} \right]  \\
\label{eq:EDPf60B} 
 \partial_r(\alpha^2\,\partial_r(r\,f_{6,0}^{B})) - \frac{42}{r} \, f_{6,0}^{B} & = - 18 \, \sqrt{\frac{35}{143}} \, \mu_0 \, \omega \, f_{5,0}^{D}
 \end{align}
\end{subequations}
The associated boundary conditions are
\begin{subequations}
\label{eq:CLfD}
\begin{align}
 \alpha^2 \, \partial_r(r\,f_{1,0}^{D}) & = 
 - \varepsilon_0 \, r \, \tilde{\omega} \, \left[ \frac{2}{\sqrt{5}} \, f_{2,0}^{B} + 2 \, \sqrt{\frac{2\,\upi}{3}} \, B \, R \right] \\
 \alpha^2 \, \left[ 2 \, \sqrt{\frac{3}{35}} \, \partial_r(r\,f_{3,0}^{D}) - \sqrt{\frac{2}{15}} \, \partial_r(r\,f_{1,0}^{D}) \right] & = 
 \varepsilon_0 \, r \, \tilde{\omega} \, \left[ \frac{10}{7} \, \sqrt{\frac{2}{3}} \, f_{2,0}^{B} - \frac{8}{7} \, f_{4,0}^{B} + \frac{4}{3} \, \sqrt{\frac{\upi}{5}} \, B \, R \right] \\
 \alpha^2 \, \left[ \sqrt{\frac{10}{33}} \, \partial_r(r\,f_{5,0}^{D}) - \frac{2}{\sqrt{21}} \, \partial_r(r\,f_{3,0}^{D}) \right] & = 
 \varepsilon_0 \, r \, \tilde{\omega} \, \left[ - \frac{4}{7} \, \sqrt{\frac{6}{5}} \, f_{2,0}^{B} + \frac{76}{77} \, \sqrt{5} \, f_{4,0}^{B} - \frac{10}{11}\,\sqrt{\frac{42}{13}} \, f_{6,0}^{B} \right]
\end{align}
\end{subequations}
where quantities have to be evaluated on the neutron star surface, at $r=R$. Details on the derivation of these equations can be found in \cite{2013MNRAS.433..986P}.
We emphasize that the magnetic field at the neutron star surface is exactly matched to the expression for the general-relativistic monopole, equation~(\ref{eq:MonopoleB}). All other multipole fields $f_{l,0}^{B}$ with $l\geq1$ vanish at $r=R$ by our definition.

\subsubsection{Numerical solution}

The above system of boundary value problems is efficiently solved by means of rational Chebyshev polynomials. The technique is presented in detail in \cite{2013MNRAS.433..986P}. Here we only report the results for the coefficients $f_{l,0}^{D}$ and $f_{l,0}^{B}$ for the monopole.

For concreteness, in all the computations, we use the following set of parameters namely $R/\Rs=\{2,2000\}$ and $\rlight/R=\{10,1000\}$ which should correspond to a compact and a non compact star as well as to a mildly rotating and a slowly rotating star.

\begin{figure*}
  \centering
  \begin{tabular}{cc}
  \includegraphics[width=0.5\textwidth]{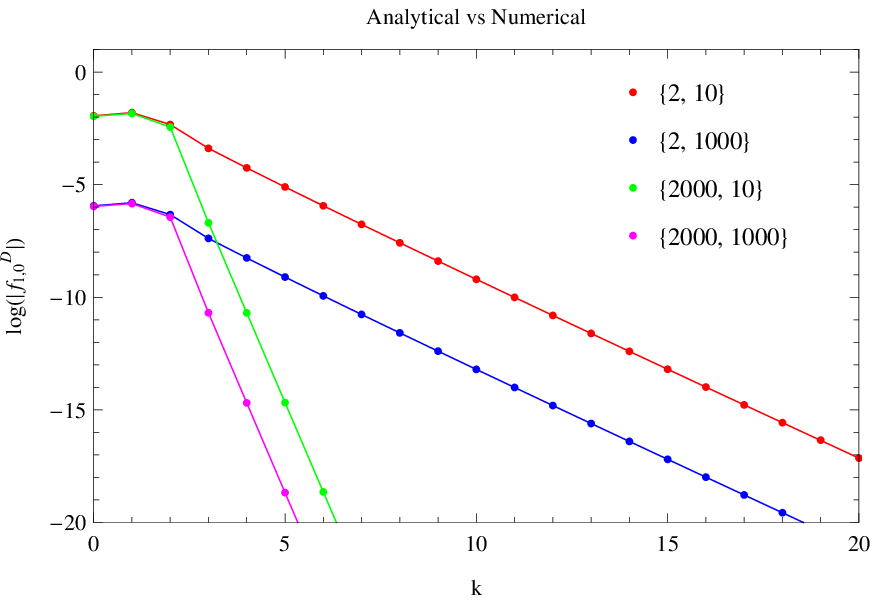} &
  \includegraphics[width=0.5\textwidth]{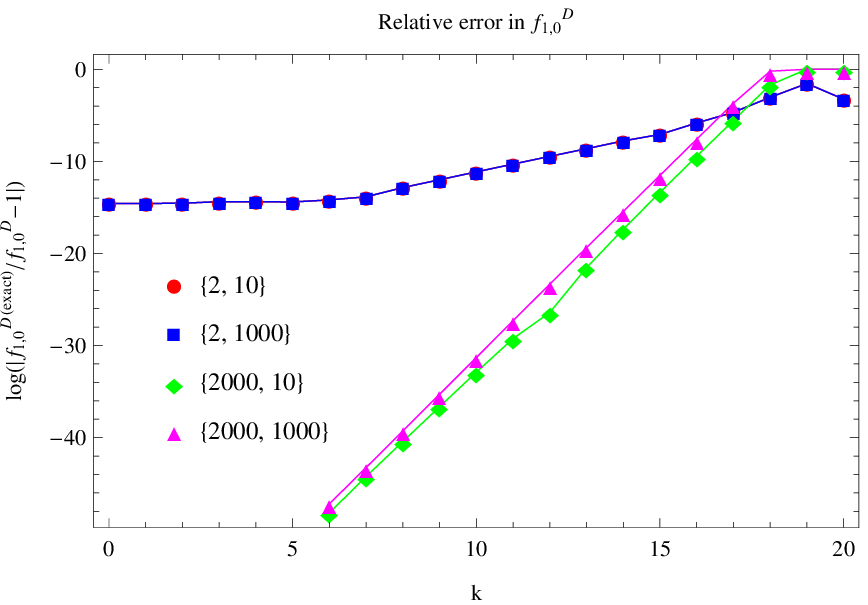}
  \end{tabular}
  \caption{Comparison of the analytical and numerical coefficients of the rational Chebyshev expansion of $f_{1,0}^D$. The absolute values of the coefficients of $f_{1,0}^D$ are shown on a logarithmic scale on the left panel and the relative error on the right panel. The solid lines correspond to the coefficients computed from the analytical exact expression whereas the dots correspond to the computed values from the boundary value problem. The inset legend shows the couple of ratios $\{R/\Rs,\rlight/R\}$.}
  \label{fig:Monopole_j0_fD10_coeff}
\end{figure*}

First we only consider the dipolar electric field component induced by the rotation of the neutron star. Strictly speaking, we should retrieve the analytical approximation equation~(\ref{eq:fD10}). This is indeed what we checked. In figure~\ref{fig:Monopole_j0_fD10_coeff} we show on the left panel the absolute value of these expansion coefficients $f_{1,0}^D$ on a logarithmic scale and on the right panel the relative error. We consider two sets, the first one computed from the analytical exact expression and the second one obtained from the numerical integration of the boundary value problem. The agreement between both solutions is excellent, the error being less than $10^{-15}$ which correspond to the double precision arithmetic of $\varepsilon=10^{-16}$. The coefficients decrease exponentially fast demonstrating the rapid convergence of the series to the exact solution. This exponential convergence to the exact solution is typical for spectral methods when the solution is $\mathcal{C}^\infty$. The relative error increases systematically when the coefficients become of the order $\varepsilon\,f_{1,0}^D$. These weak coefficients cannot be computed accurately because of the finite precision of the computer. This is of no concern as in any expansion series, they become irrelevant because not contributing to the summation in a significant way.

\begin{figure*}
  \begin{tabular}{cc}
  \includegraphics[width=0.5\textwidth]{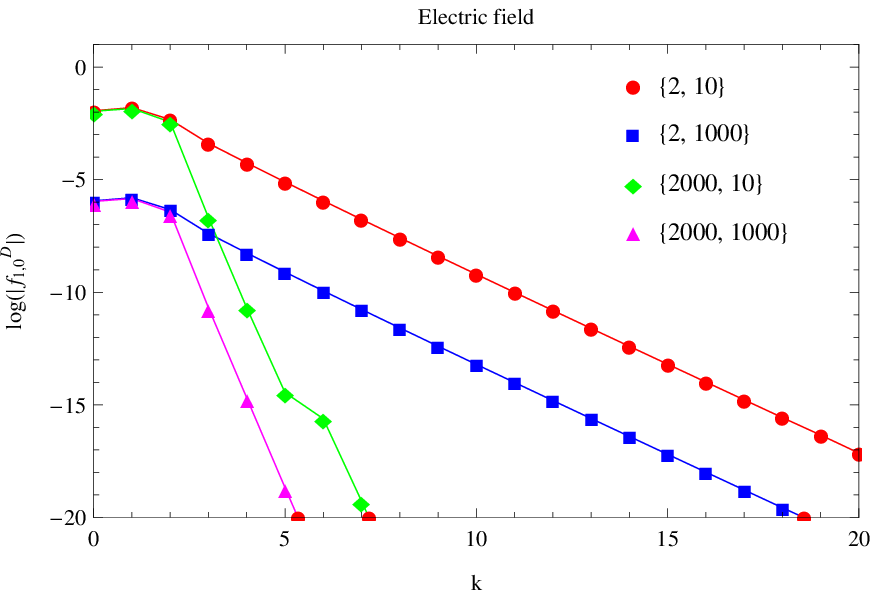} &
  \includegraphics[width=0.5\textwidth]{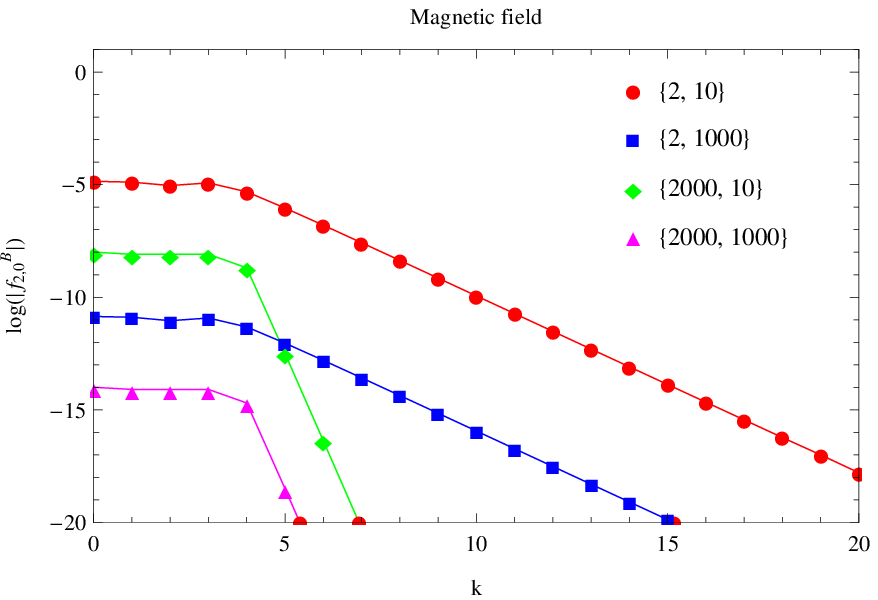}
  \end{tabular}
  \caption{Coefficients of the rational Chebyshev expansion of $f_{1,0}^D$ and $f_{2,0}^B$. Their absolute values are shown on a logarithmic scale. The inset legend shows the couple of ratios $\{R/\Rs,\rlight/R\}$.}
  \label{fig:Monopole_j0_fD10_fB20_coeff}
\end{figure*}

\begin{figure*}
  \begin{tabular}{cc}
  \includegraphics[width=0.5\textwidth]{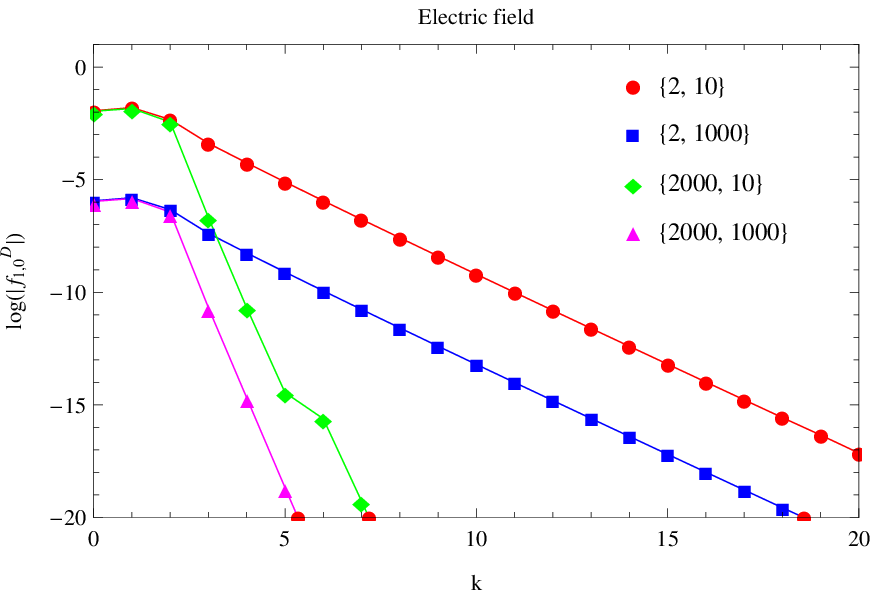} &
  \includegraphics[width=0.5\textwidth]{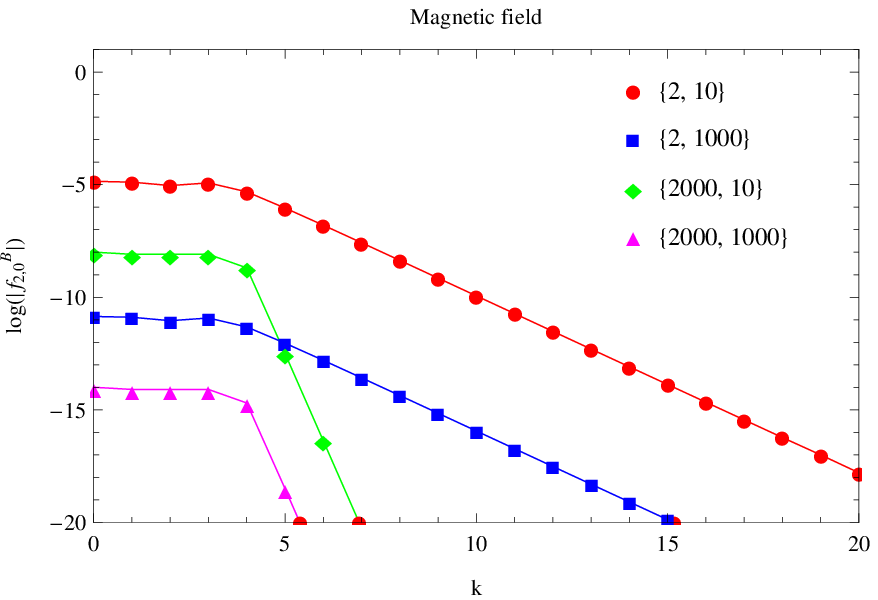} \\
  \includegraphics[width=0.5\textwidth]{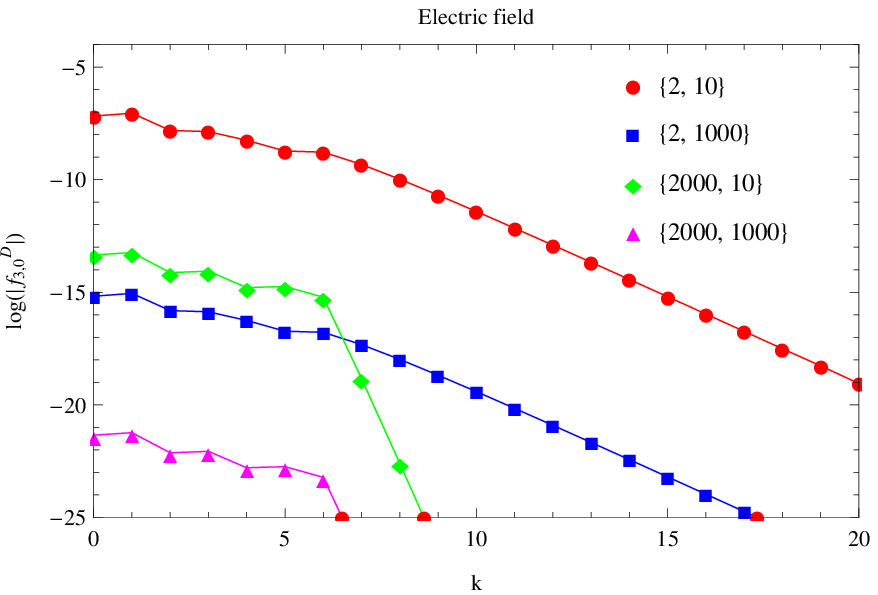} &
  \includegraphics[width=0.5\textwidth]{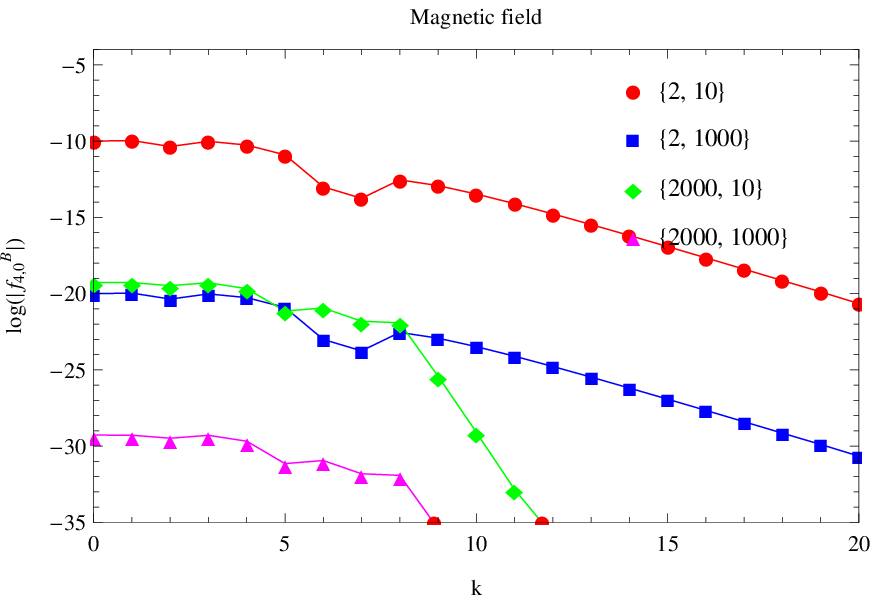}
  \end{tabular}
  \caption{Coefficients of the rational Chebyshev expansion of $f_{1,0}^D$, $f_{2,0}^B$, $f_{3,0}^D$ and $f_{4,0}^B$. Their absolute values are shown on a logarithmic scale. The inset legend shows the couple of ratios $\{R/\Rs,\rlight/R\}$. Note the different scales used for each plot.}
  \label{fig:Monopole_j0_fD10_fB20_fD30_fB40_coeff}
\end{figure*}

\begin{figure*}
  \begin{tabular}{cc}
  \includegraphics[width=0.5\textwidth]{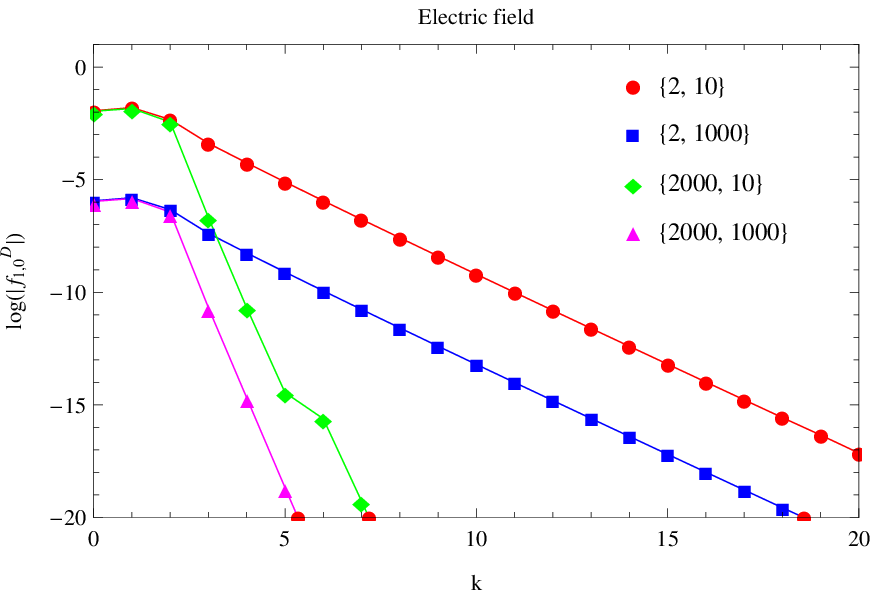} &
  \includegraphics[width=0.5\textwidth]{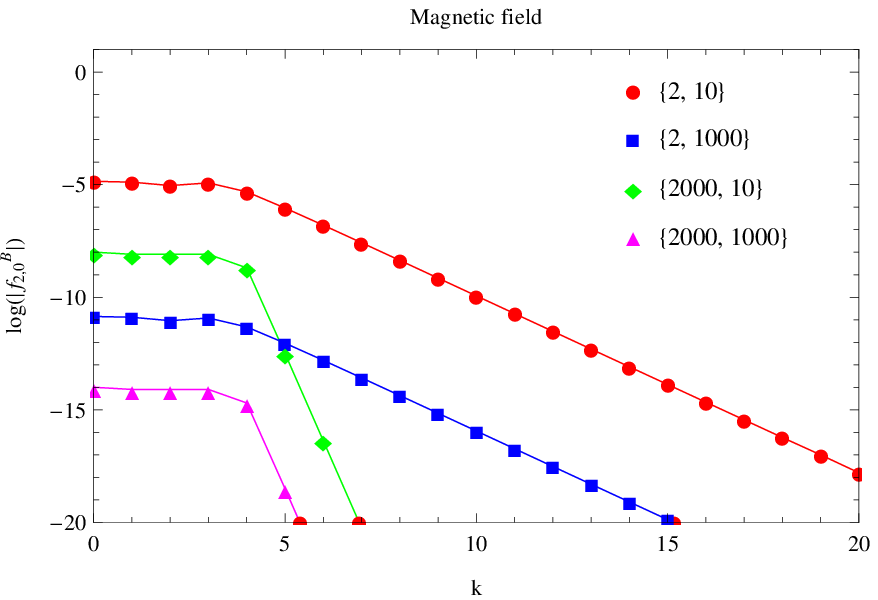} \\
  \includegraphics[width=0.5\textwidth]{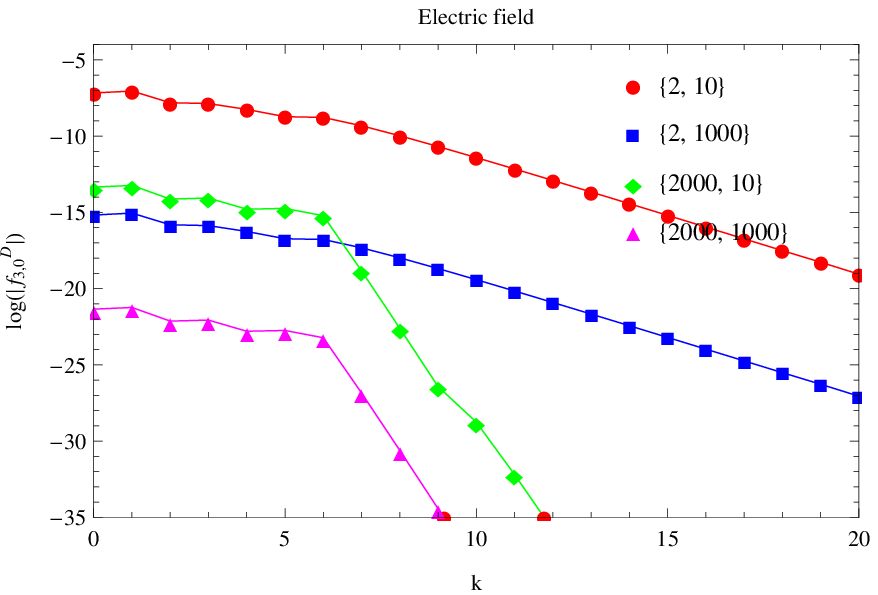} &
  \includegraphics[width=0.5\textwidth]{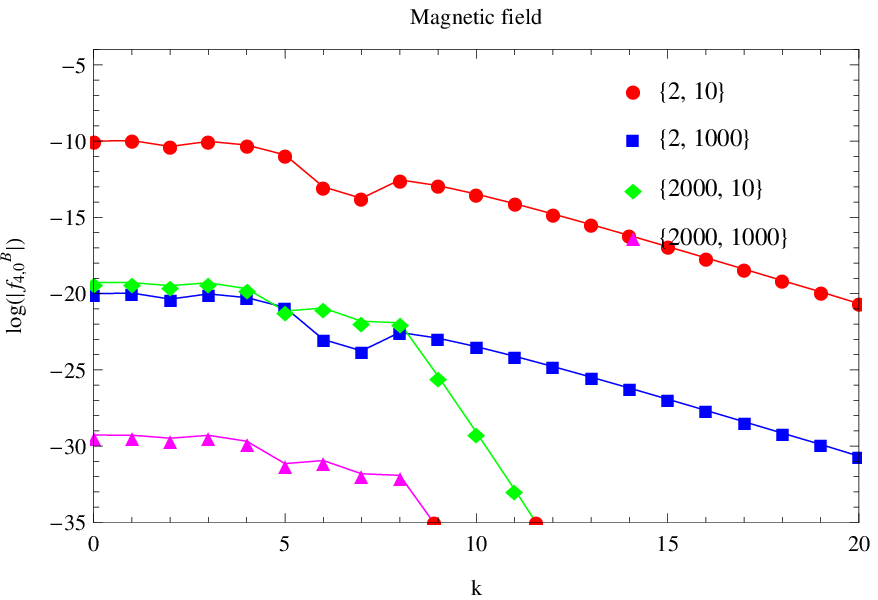} \\
  \includegraphics[width=0.5\textwidth]{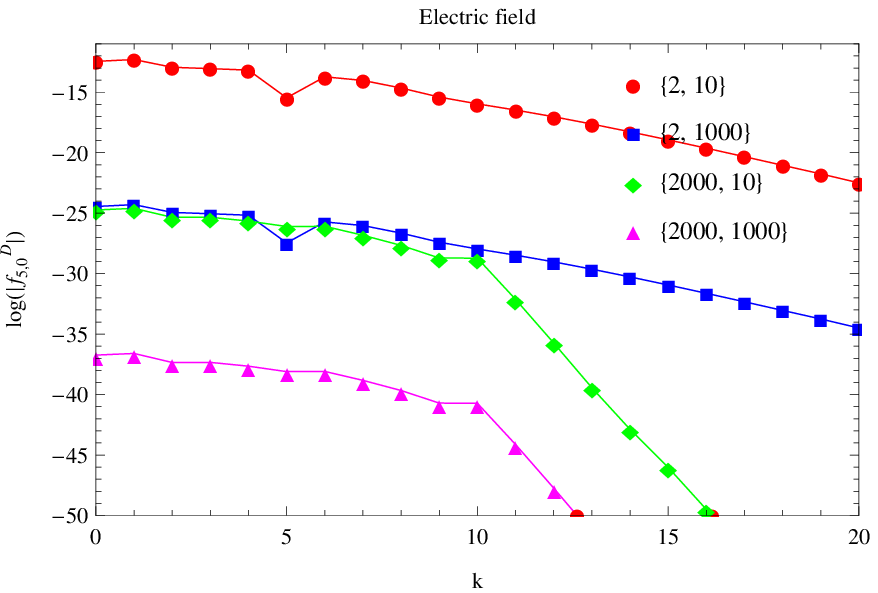} &
  \includegraphics[width=0.5\textwidth]{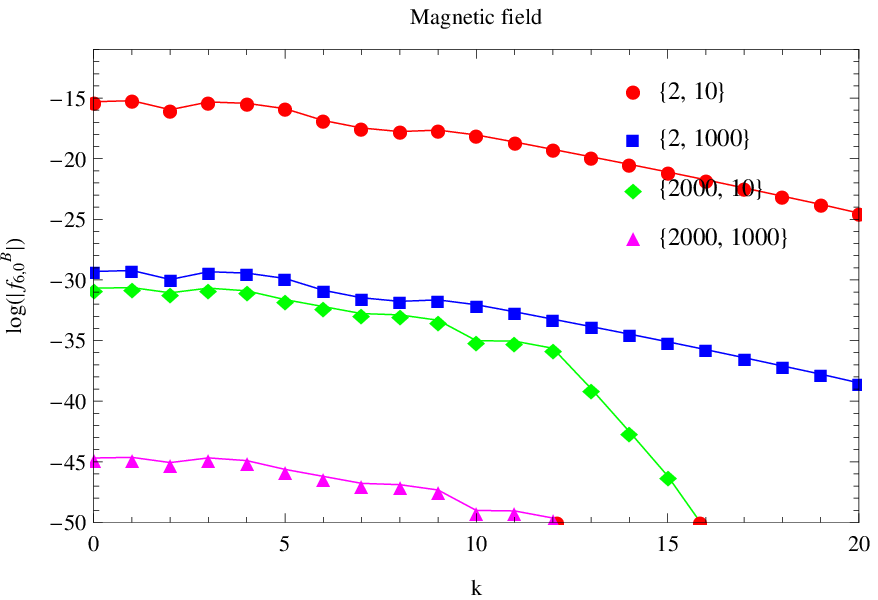}
  \end{tabular}
  \caption{Coefficients of the rational Chebyshev expansion of $f_{1,0}^D$, $f_{2,0}^B$, $f_{3,0}^D$, $f_{4,0}^B$, $f_{5,0}^D$ and $f_{6,0}^B$. Their absolute values are shown on a logarithmic scale. The inset legend shows the couple of ratios $\{R/\Rs,\rlight/R\}$.  Note the different scales used for each plot.}
  \label{fig:Monopole_j0_fD10_fB20_fD30_fB40_fD50_fB60_coeff}
\end{figure*}

After this first test of the solution to the boundary value problem, we switch to the next order of approximation including a perturbation in the magnetic field which will be of quadrupolar order. We thus have to solve simultaneously for $f_{1,0}^D$ and $f_{2,0}^B$. In order to show the rapid convergence of the coefficients, we plot again their absolute values in logarithmic scale, as depicted in figure~\ref{fig:Monopole_j0_fD10_fB20_coeff}.
For the next approximation, we add the multipolar coefficients $f_{3,0}^D$ and $f_{4,0}^B$.
Convergence is proven by inspection of figure~\ref{fig:Monopole_j0_fD10_fB20_fD30_fB40_coeff} showing an exponential decay of the coefficients with respect to the index $k$.
Finally, for the most accurate solution we put two other multipolar components, namely $f_{5,0}^D$ and $f_{6,0}^B$.
Figure~\ref{fig:Monopole_j0_fD10_fB20_fD30_fB40_fD50_fB60_coeff} compares the relative importance of each multipolar component with respect to each other. We always observe the characteristic exponential convergence as expected in this smooth boundary value problem.
All the coefficients of the electric and magnetic field functions decrease exponentially fast. The rational Chebyshev expansion is very effective in approximating the true analytical solution with only a few terms.

\begin{figure*}
  \begin{tabular}{cc}
  \includegraphics[width=0.5\textwidth]{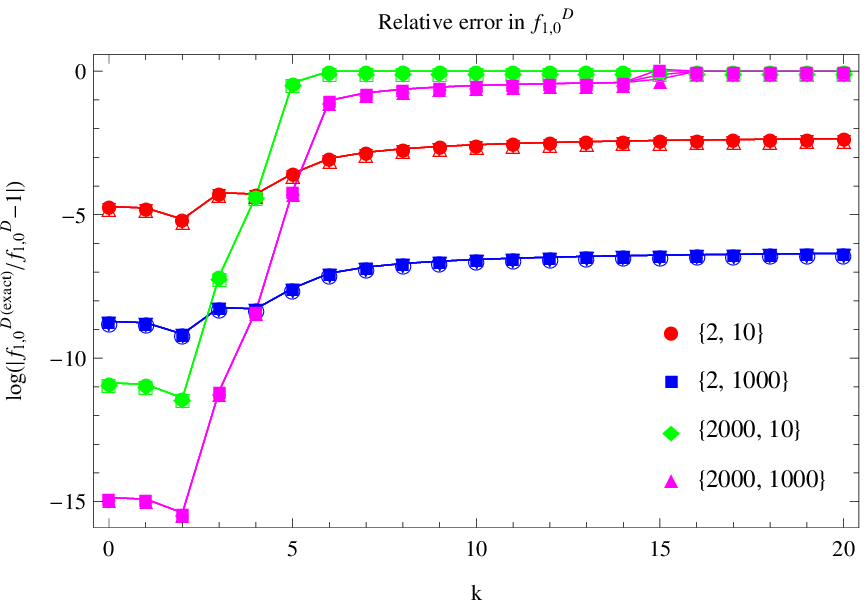} &
  \includegraphics[width=0.5\textwidth]{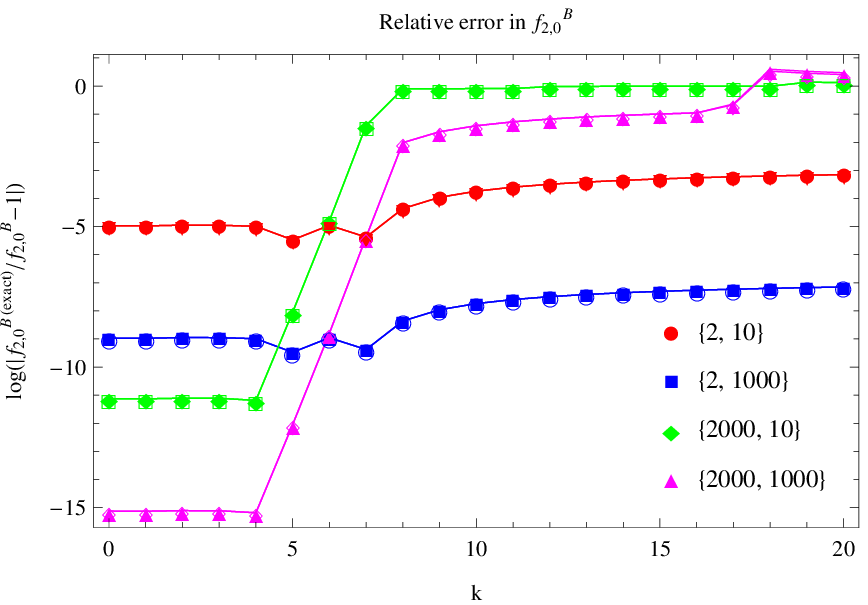}
  \end{tabular}
  \caption{Relative error of the rational Chebyshev expansion of $f_{1,0}^D$ and $f_{2,0}^B$ when adding multipolar components. The inset legend shows the couple of ratios $\{R/\Rs,\rlight/R\}$.  For a given couple $\{R/\Rs,\rlight/R\}$, all curves almost overlap whatever the number of multipoles added into the expansion.}
  \label{fig:Erreur_Monopole_j0_fD10_fB20_coeff}
\end{figure*}
The relative errors of the rational Chebyshev expansion of $f_{1,0}^D$ and $f_{2,0}^B$ when adding multipolar components are shown in fig.~\ref{fig:Erreur_Monopole_j0_fD10_fB20_coeff}. Comparing to the strongest perturbation induced by the presence of only $\{f_{3,0}^D, f_{4,0}^B\}$ the difference remains insensitive when the modes $\{f_{5,0}^D, f_{6,0}^B\}$ are present. Adding higher mulitpole components to the expansion series will not drastically change the lowest order coefficients $f_{1,0}^D$ and $f_{2,0}^B$, at most only starting from the fifth digit. Indeed, for a given couple $\{R/\Rs,\rlight/R\}$, all curves almost overlap whatever the number of multipoles added into the expansion. Multipolar fields higher than $l=4$, although present are definitely too weak to have an influence on the electric dipole and magnetic quadrupole fields.

\begin{figure*}
\begin{tabular}{cc}
  \includegraphics[width=0.5\textwidth]{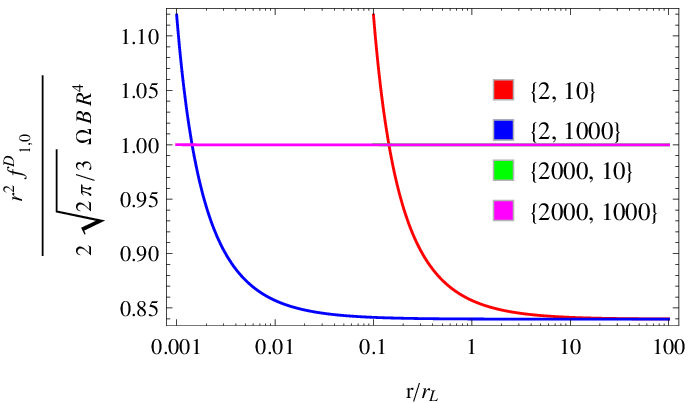} &
  \includegraphics[width=0.5\textwidth]{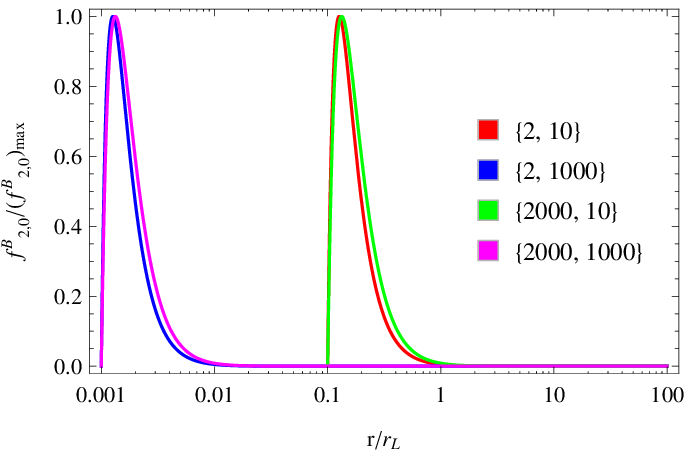} \\
  \includegraphics[width=0.5\textwidth]{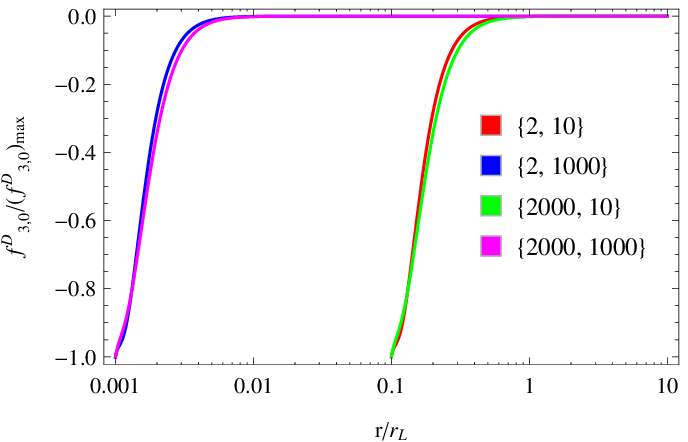} &
  \includegraphics[width=0.5\textwidth]{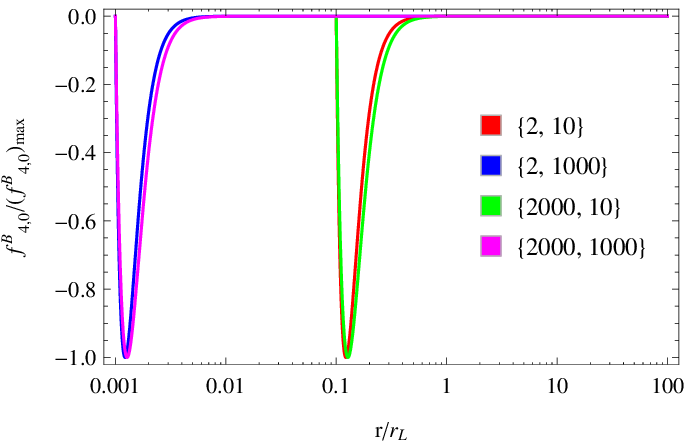} \\
  \includegraphics[width=0.5\textwidth]{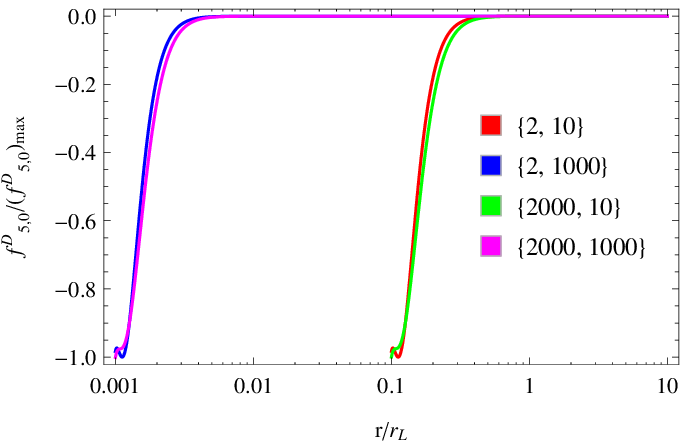} &
  \includegraphics[width=0.5\textwidth]{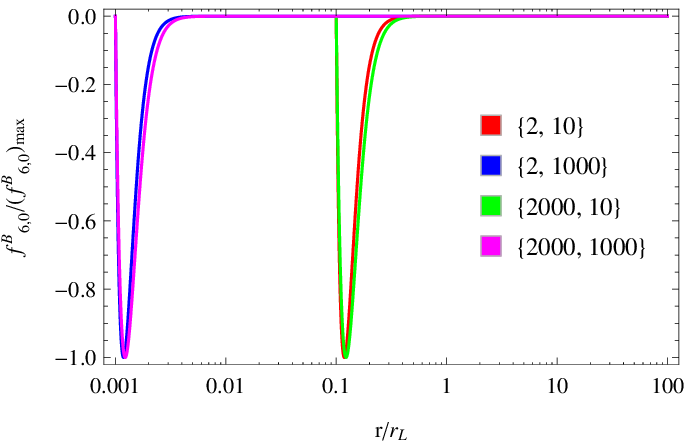}
\end{tabular}
  \caption{The radial profile of the electric and magnetic functions $\{f_{1,0}^D, f_{3,0}^D, f_{5,0}^D\}$ and $\{f_{2,0}^B, f_{4,0}^B, f_{6,0}^B\}$. For convenience, the functions are normalized except for $f_{1,0}^D$. The inset legend indicates the couple of ratio $\{R/\Rs,\rlight/R\}$.}
  \label{fig:Monopole_j0_fD10_fB20_fD30_fB40_fD50_fB60_rad}
\end{figure*}

To conclude this section, we plot the radial dependence of the functions $\{f_{1,0}^D, f_{2,0}^B, f_{3,0}^D, f_{4,0}^B, f_{5,0}^D, f_{6,0}^B\}$ in the {four cases corresponding to a slowly or rapidly rotating star, compact or not, with parameters $R/\Rs=\{2,2000\}$ and $\rlight/R=\{10,1000\}$, see figure~\ref{fig:Monopole_j0_fD10_fB20_fD30_fB40_fD50_fB60_rad}}. These functions can then directly be compared to the output of our numerical simulations in section~\ref{sec:Results}. The functions are normalized in order to put them on a same graph except for $f_{1,0}^D$ which is the leading term.
 The non compact object case with $R/\Rs=2000$ remains very close to the flat space-time solution. Thus a good approximation to the electric field is given by equation~(\ref{eq:f10E}). This is clearly seen in the upper left panel of figure~\ref{fig:Monopole_j0_fD10_fB20_fD30_fB40_fD50_fB60_rad}.
In principle, we are able to compute the electromagnetic field to any order to get the solution to any required precision. Actually we stopped with a three terms expansion in the electric and magnetic field respectively because high order multipole moments become negligible compared to the lowest order. We also think that it is largely enough to compare with the numerical code we now describe in details.

\twocolumn

\section{CODE DESCRIPTION}
\label{sec:Algorithme}

We now give the general outline of our pseudo-spectral discontinuous Galerkin finite element algorithm. The main ingredients are, the expansion on to vector spherical harmonics for divergencelessness fields in spherical shells, an exact imposition of boundary conditions on the neutron star surface, an explicit time stepping with a fourth-order Runge-Kutta integration scheme, a spectral filtering in the longitudinal and latitudinal directions and a limiting procedure in the radial direction. The radial part is solved with a high-order finite volume scheme whereas the spherical part is solved through a pseudo-spectral approach.

\subsection{One dimensional scalar conservation law}

To present our new code, we will focus on the one dimensional scalar conservation law which is an archetypal of equations often used to model physical phenomena. Consider therefore the simple conservation law of a scalar field denoted by~$u$ with a physical flux function denoted by~$f$ such that the conservation of~$u$ is expressed as a partial differential equation written as
\begin{equation}
\label{eq:Conservation}
 \partial_t u + \partial_x f(u) = 0.
\end{equation}
This equation has to be solved for any time $t\geq0$ and for all $x\in[a,b]$ where $[a,b]$ is the computational domain. Note that in our code $x$ should be interpreted as the radial coordinate~$r$. We subdivide the domain~$[a,b]$ in $K$~cells not necessarily of the same length. In each of these cells which we denote by~$D^k$ with $k\in[0..K-1]$, the solution is expanded on to a basis of spatial functions~$\phi^k_i$ such that the approximate solution in the cell~$k$ reads
\begin{equation}
 u^k(x,t) = \sum_{i=0}^{N_p} u^k_i(t) \, \phi^k_i(x)
\end{equation}
valid in the cell~$k$ given by the interval~$[x^k_l,x^k_r]$. The basis possesses $N_p+1$ functions. The spatial method is therefore of order~$N_p$. After injecting this expansion into the conservation law equation~(\ref{eq:Conservation}) and projecting on to the basis functions~$\phi_i^k$, performing two successive integrations by part in each cell independently, starting from
\begin{equation}
 \int_{x^k_l}^{x^k_r} ( \partial_t u + \partial_x f(u) ) \, \phi_i^k \, dx = 0
\end{equation}
we arrive at the strong form of the partial differential equation such that
\begin{equation}
 \sum_{j=0}^{N_p} ( \int_{x^k_l}^{x^k_r} l_i^k \, l_j^k \, dx ) \, \partial_t u_j^k + \sum_{j=0}^{N_p} ( \int_{x^k_l}^{x^k_r} l_i^k \, \partial_x l_j^k \, dx) \, f_j^k = [ ( f - f^*) \,l_i^k ]_{x^k_l}^{x^k_r}
\end{equation}
We introduced a numerical flux $f^*$ which tells to the system how to communicate information between adjacent cells as in classical finite volume schemes. Taking for instance $\phi_i^k(x)=\tilde{P}_i(y_k(x))$ which are the normalized Legendre polynomials defined on the interval~$[-1,1]$ and $y_k(x)$ a scaling function to shift from $x\in[x^k_l,x^k_r]$ to $y_k\in[-1,1]$, we introduce the following matrices
\begin{subequations}
\begin{align}
 \mathcal{M}_{ij}^k & = \int_{x^k_l}^{x^k_r} \phi_i^k \, \phi_j^k \, dx \\
 \mathcal{S}_{ij}^k & = \int_{x^k_l}^{x^k_r} \phi_i^k \, \partial_x \phi_j^k \, dx
\end{align}
\end{subequations}
These matrices can be computed analytically and exactly. The semi-discrete system to be solved then becomes
\begin{equation}
 \label{eq:GalerkinDiscontinue}
 \sum_{j=0}^{N_p} \mathcal{M}_{ij}^k \, \partial_t u_j^k + \sum_{j=0}^{N_p} \mathcal{S}_{ij}^k \, f_j^k = [ ( f - f^*) \, \tilde{P}_i^k ]_{x^k_l}^{x^k_r}
\end{equation}
or in a pure matrix notation
\begin{equation}
 \label{eq:GalerkinDiscontinueMat}
 \mathcal{M}^k \, \partial_t \mathcal{U}^k + \mathcal{S}^k \, \mathcal{F}^k = [ ( f - f^*) \, \mathcal{P}^k ]_{x^k_l}^{x^k_r}
\end{equation}
with $\mathcal{P}^k$ the column vector of the normalized Legendre polynomials. Note that for an orthogonal basis, the mass matrix is diagonal hence very easy to invert. Inverting the mass matrix~$\mathcal{M}^k$, each coefficient of the expansion evolves according to the first order ordinary differential equation
\begin{equation}
 \label{eq:GalerkinDiscontinueMatricielle}
 \partial_t \mathcal{U}^k + (\mathcal{M}^k)^{-1} \, \mathcal{S}^k \, \mathcal{F}^k = (\mathcal{M}^k)^{-1} \, [ ( f - f^*) \, \mathcal{P}^k ]_{x^k_l}^{x^k_r}
\end{equation}
The state of the art in the discontinuous Galerkin methods resides in the choice of the numerical flux~$f^*$ which has to satisfy several stability and consistence properties. The reader is referred to the excellent book by \cite{Hesthaven2008} for a detailed discussion about the implementation of modal and nodal discontinuous Galerkin methods in one dimension and the tricks to deal with non-linear problems, introducing limiting and filtering processes. Here we only give guide lines on the way to implement the techniques for spherical geometries. Let us first discuss the main advantage of the code, namely the flexibility in the choice of the grid.

\subsection{The grid}

Our goal is to look deeply into the light-cylinder with very small ratios of neutron star radius to light-cylinder radius, $R/\rlight\ll1$, as well as far away from the light-cylinder at distances $r$ much larger than $\rlight$, $r/\rlight\gg1$. In our previous work \cite{2012MNRAS.424..605P}, we had some difficulties to achieve such demanding parameters because we used only one radial domain to expand on to Chebyshev polynomials. We thought that the code could greatly benefit from a more advantageous domain decomposition in the radial direction. Indeed, this allows us to zoom into the light-cylinder with very fine grids close to the surface but keeping a coarser grid outside the light-cylinder where we can afford a loss in precision for sufficiently large radii. Due to the flexibility of domain decomposition methods, we are able to use a non-uniform grid when moving from one radial cell to the next one. This technique is called spectral element method \citep{Canuto2007}. It can be seen as a high-order finite volume scheme. To use all the advantages of the conservative form of such finite volume formulation, we prefer to expand the radial direction into normalized Legendre polynomials instead of Chebyshev polynomials. Such expansion makes the algorithm rigorously conservative, meaning that the average value of the unknown quantities are perfectly conserved during the simulation, within numerical round-off errors. 

The arbitrary nature of the radial scale is used to fix small volumes close to the neutron star whereas larger shells are sufficient farther away. To be more specific, we employ the usual Fourier transform in the $\{\vartheta,\varphi\}$ directions and expand the radial coordinate into $K$ sub-intervals (which can be seen as finite volume elements), the boundary of each cell is given by~$[r_g^k,r_d^k]$ with $k\in[0..K-1]$ dividing the global interval $[R_1,R_2]$ into non necessarily equal sub-intervals. In each of these volumes, we expand the radial part into normalized Legendre polynomials by rescaling each interval $[r_g^k,r_d^k]$ into $[-1,1]$ through a scaling function. 

Let us assume that the computational domain is comprised between the neutron star surface at~$R_1=R$ and an arbitrary outer radius~$R_2$. The spherical shell is decomposed into $K$ cells but with increasing thickness. We introduce two temporary variables $y_1=\log(R_1/\rlight)$ and $y_2=\log(R_2/\rlight)$ and a logarithmic thickness by $h=(y_2-y_1)/K$. Each cell, labelled with a superscript~$k$, possesses then two interfaces located at
\begin{subequations}
 \begin{align}
  r_g^k & = e^{y_1+k\,h} \\
  r_d^k & = e^{y_1+(k+1)\,h}
 \end{align}
\end{subequations}
The thickness of the cell labelled~$k$ is $h^k=r_d^k-r_g^k$.
In that way, the ratio between the size of two successive cells is constant and equal to $e^h$. We will show that such variable cell size drastically improves the accuracy in the innermost parts of the simulation box while preserving good accuracy well outside the light-cylinder.

\subsection{Vector expansion and divergencelessness constraint on $\bmath{B}$}

We use again a clever expansion of the vector fields $\bmath{B}$ and $\bmath{D}$. Indeed, electric and magnetic fields are expanded onto vector spherical harmonics (VSH) according to
\begin{subequations}
 \begin{align}
   \label{eq:D_vhs}
  \bmath{D} & = \sum_{l=0}^\infty\sum_{m=-l}^l \left(D^r_{lm} \, \bmath{Y}_{lm} + D^{(1)}_{lm} \, \bmath{\Psi}_{lm} + D^{(2)}_{lm} \, \bmath{\Phi}_{lm}\right) \\
  \label{eq:B_vhs}
  \bmath{B} & = \sum_{l=0}^\infty\sum_{m=-l}^l \left(B^r_{lm} \, \bmath{Y}_{lm} + B^{(1)}_{lm} \, \bmath{\Psi}_{lm} + B^{(2)}_{lm} \, \bmath{\Phi}_{lm}\right)
\end{align}
\end{subequations}
Such expansion is done in each cell. However, in order to deal with the divergencelessness of the magnetic field whatever the configuration of the electromagnetic field, loaded or not with plasma it is more appropriate to use an expansion of $\bmath{B}$ into
\begin{equation}
  \label{eq:B_div0}
  \bmath{B} = \sum_{l=1}^\infty\sum_{m=-l}^l \rot [f^B_{lm}(r,t) \, \bmath{\Phi}_{lm}] + g^B_{lm}(r,t) \, \bmath{\Phi}_{lm} 
\end{equation}
where $\{f^B_{lm}(r,t), g^B_{lm}(r,t)\}$ are the expansion coefficients of $\bmath{B}$. The monopole part, eq.~(\ref{eq:MonopoleB}), is added by hand. To impose the divergencelessness constraint, we project the magnetic field on to the subspace subtended by the expansion in equation~(\ref{eq:B_div0}). Actually, because spectral methods for smooth problems are very accurate, the projection is not required at each time step. We perform it only when the divergence becomes larger than a threshold defined by the user.

\subsection{Numerical flux}

As in any other finite volume scheme, communication between cells goes through a numerical flux~$f^*$ chosen to resolve as accurately as possibly the conservation laws. In the force-free limit, the dynamics reduce to the solution of Maxwell equations with source terms. So we only need to find an appropriate numerical flux for the linear advection problem in one dimension, namely the radial direction. The efficiency of the numerical code will strongly depend on the choice of the numerical flux. For Maxwell equations, we employ a first order upwind scheme as described in \cite{Hesthaven2008}. Starting from the 3+1~formalism, we consider the one dimensional system of Maxwell equations in spherical geometry and relevant for propagation in the radial direction. Thus only the components $(E^\vartheta, E^\varphi, H^\vartheta, H^\varphi)$ are meaningful. In this way we get the following equations describing the propagation of the electromagnetic field in the radial direction in general relativity by
\begin{subequations}
 \begin{align}
  \partial_t D^\vartheta + \frac{\alpha}{r} \, \partial_r (r\,H^\varphi) & = 0 \\
  \partial_t D^\varphi - \frac{\alpha}{r} \, \partial_r (r\,H^\vartheta) & = 0 \\
  \partial_t B^\vartheta - \frac{\alpha}{r} \, \partial_r (r\,E^\varphi) & = 0 \\
  \partial_t B^\varphi + \frac{\alpha}{r} \, \partial_r (r\,E^\vartheta) & = 0
 \end{align}
\end{subequations}
By a change of variables through the quantity
\begin{equation}
 u_{(D/B)}^{(\vartheta/\varphi)} = \frac{r \, (D/B)^{(\vartheta/\varphi)}}{\alpha}
\end{equation}
the above system becomes strictly conservative, assuming that the lapse function is time-independent. We can then apply standard discontinuous Galerkin methods to our problem. Introducing the jumps of the electromagnetic field components at the cell interface, denoted by~$du=u_d-u_g$, the associated numerical upwind flux becomes
\begin{equation}
 f^* = \frac{r}{2}
\begin{pmatrix}
  H_d^\varphi + H_g^\varphi + \frac{D_g^\vartheta-D_d^\vartheta}{\alpha} \\
 -(H_d^\vartheta + H_g^\vartheta) + \frac{D_g^\varphi-D_d^\varphi}{\alpha} \\
 -(E_d^\varphi + E_g^\varphi) + \frac{B_g^\vartheta-B_d^\vartheta}{\alpha} \\
  E_d^\vartheta + E_g^\vartheta + \frac{B_g^\varphi-B_d^\varphi}{\alpha}  
\end{pmatrix}
\end{equation}
From these expressions, we deduce the right hand side on the left interface of a cell by
\begin{equation}
 f_d - f^* = \frac{r}{2\,\alpha}
\begin{pmatrix}
  dD^\vartheta + \alpha \, dH^\varphi \\
  dD^\varphi - \alpha \, dH^\vartheta \\
  dB^\vartheta - \alpha \, dE^\varphi \\
  dB^\varphi + \alpha \, dE^\vartheta \\
\end{pmatrix}
\end{equation}
and the corresponding right hand side on the right interface of a cell by
\begin{equation}
 f_g - f^* = \frac{r}{2\,\alpha}
\begin{pmatrix}
  dD^\vartheta - \alpha \, dH^\varphi \\
  dD^\varphi + \alpha \, dH^\vartheta \\
  dB^\vartheta + \alpha \, dE^\varphi \\
  dB^\varphi - \alpha \, dE^\vartheta \\
\end{pmatrix}
\end{equation}
These numerical fluxes close the overall description of the basic algorithm.
We now switch to the delicate problem of non-linearities and how to overcome aliasing effects and related numerical instabilities.

\subsection{Slope Limiter}

The slope limiting technique is adapted from the classical finite volume community. The idea is to reduce or even kill spurious oscillations that arise from the non-linear evolution or from sharp discontinuities in the solution. The most basic total variation diminishing (TVD) limiters are usually too dissipative for higher-order schemes. \cite{Toro2009} detailed several TVD schemes with application to simple problems and compares the merit of each slope limiter. We refer the reader to this book for more information about the use of TVD method in finite volume algorithms. Indeed, in trouble cells, the polynomial expansion is reduced to at most a linear interpolation and therefore considerably reducing the order of the method around discontinuities. To circumvent such drawbacks, it is necessary to release the TVD property for a less stringent property called total variation bound (TVB) method \citep{1989JCoPh..84...90C}. The latter does not guaranty strict cancellation of oscillations but only weaken them whereas the former completely avoids oscillations but at the cost of reducing to a low-order scheme. In the simulations shown in this paper, we found that the TVB limiter represents a good compromise between accuracy and spurious oscillations. We implemented both limiters and checked that TVB is preferable to TVD limiters. Unfortunately TVB methods introduce one more parameter, often depicted by the capital letter~$M$. Moreover the value of this parameter is very problem dependent, related to the second spatial derivative of the solution, therefore a priori unknown. So we let the user arbitrarily choose the best limiter parameter~$M$ by some trial and error tests. Various examples of limiters can be found in the literature, see \cite{Hesthaven2008} for some basic discussion, including the difference between TVD and TVB. In our algorithm, we tried the MUSCL limiter 
and the less dissipative TVBM limiter. For high enough resolution we did not find any significant difference between both limiters. Thus we will not discuss the influence of these limiters on the solution.

\subsection{Filtering}

The limiter cannot be applied in the latitudinal and longitudinal direction simply because there is no domain decomposition in those directions. We use the classical spherical harmonic expansion. The force-free problem being non-linear due to the electric current in the source terms, we expect the solution to develop sharp gradients or discontinuities also in the spherical directions. It is therefore compulsory to get rid of these high frequencies by some filtering procedure. This is achieved by adding a small damping factor to the high order coefficients of the expansion in $Y_{lm}$. Filtering is performed at each time step. We use an exponential filter in directions~$(\vartheta,\varphi)$ given by the general expression
\begin{equation}
  \label{eq:Filtre}
  \sigma(\eta) = \textrm{e}^{-\alpha\,\eta^\beta}
\end{equation}
where the variable~$\eta$ ranges between 0 and 1. For instance, in the latitudinal direction $\eta = l/(N_{\vartheta}-1)$ for $l\in[0..N_{\vartheta}-1]$, $l$ being the index of the coefficient $c_{l,m}$ in the spherical harmonic expansion $f(\vartheta,\varphi) = \sum_{l,m=0}^{N_{\vartheta}-1,N_{\varphi}-1} c_{l,m} \,Y_{l,m}(\vartheta,\varphi)$ and $N_{\vartheta},N_\varphi$ the number of collocation points in the spherical direction (latitude and longitude). The parameter $\alpha$ (not to be confused with the lapse function) is adjusted to values not too large in order to avoid errors in the solution but also not too small in order to sufficiently damp these oscillations. 

The above mentioned exponential filter of order~$\beta$ does not strictly satisfy the condition for the smoothing factors as explained in \cite{Canuto2006}. However, for numerical purposes we choose $\alpha$ such that $\textrm{e}^{-\alpha}$ is numerically zero i.e. below the machine accuracy $\varepsilon$. In practice, we choose $\alpha=36$ assuming double precision computation with $\epsilon\approx10^{-15}$. The order~$\beta$ of the smoothing influences the dissipation rate in the solution. The low order multipole components are weakly damped and correspond to large scale structures. If the solution shows fine scale structures, the filtering has to be minimized. We will discuss the role of $\beta$ in the particular case of the split monopole solution in the next sections. We typically tried $\beta\in\{2,4,8\}$. Actually, because higher order multipoles are almost absent in the solutions, let it be vacuum or force-free, a low order filtering was enough to reach satisfactory accuracy. In all the simulations presented in this work, if not explicitly specified, we systematically used a fourth order filter with $\beta=4$. We also tried a second and eighth order filter without significant variation in the solution. The split monopole is a notable exception for which higher order filtering and a large number of collocation points are necessary to correctly catch the discontinuity induced by the equatorial current sheet.

\subsection{Exact boundary conditions}

As in \cite{2014MNRAS.439.1071P} we put exact boundary conditions on the star. In general relativity the correct jump conditions at the stellar surface, continuity of the normal component of the magnetic field~$B^{\hat r}$ and continuity of the tangential component of the electric field~$\{D^{\hat \vartheta}, D^{\hat \varphi}\}$ are such that
\begin{subequations}
  \label{eq:CLimites}
\begin{align}
  B^{\hat r}(t,R,\vartheta,\varphi) & = B^{\hat r}_0(t,\vartheta,\varphi) \\
  D^{\hat \vartheta}(t,R,\vartheta,\varphi) & = - \varepsilon_0 \, \frac{\Omega-\omega}{\alpha} \, R \, \sin\vartheta \, B^{\hat r}_0(t,\vartheta,\varphi) \\
  D^{\hat \varphi}(t,R,\vartheta,\varphi) & = 0
\end{align}
\end{subequations}
The continuity of $B^{\hat r}$ automatically implies the correct boundary treatment of the electric field. $B^{\hat r}_0(t,\vartheta,\varphi)$ represents the, possibly time-dependent, radial magnetic field imposed by the star, let it be monopole, split monopole, oblique dipole or multipole.

The outer boundary condition cannot be handled exactly. We need to make some approximate assumptions about the outgoing waves we want to enforce in order to prevent reflections from this artificial outer boundary. Using the Characteristic Compatibility Method (CCM) described in \cite{Canuto2007} and neglecting the frame-dragging effect far from the neutron star, the radially propagating characteristics are given to good accuracy by 
\begin{eqnarray}
  \label{eq:CCM1}
  D^{\hat \vartheta} \pm \varepsilon_0 \, c\, B^{\hat \varphi} & ; & D^{\hat \varphi} \pm \varepsilon_0 \, c\, B^{\hat \vartheta}
\end{eqnarray}
In order to forbid ingoing wave we ensure that
\begin{subequations}
\begin{align}
  \label{eq:CCM2}
  D^{\hat \vartheta} - \varepsilon_0 \, c\, B^{\hat \varphi} & = 0 \\
  \label{eq:CCM3}
  D^{\hat \varphi} + \varepsilon_0 \, c\, B^{\hat \vartheta} & = 0
\end{align}
\end{subequations}
whereas the other two characteristics are found by
\begin{subequations}
\begin{align}
  \label{eq:CCM4}
  D^{\hat \vartheta} + \varepsilon_0 \, c\, B^{\hat \varphi} & = D^{\hat \vartheta}_{\rm PDE} + \varepsilon_0 \, c\, B^{\hat \varphi}_{\rm PDE} \\
  \label{eq:CCM5}
  D^{\hat \varphi} - \varepsilon_0 \, c\, B^{\hat \vartheta} & = D^{\hat \varphi}_{\rm PDE} - \varepsilon_0 \, c\, B^{\hat \vartheta}_{\rm PDE}
\end{align}
\end{subequations}
the subscript $_{\rm PDE}$ denoting the values of the electromagnetic field obtained by straightforward time advancing without care of any boundary condition. The new corrected values are deduced from the solution of the linear system made of equations~(\ref{eq:CCM2})-(\ref{eq:CCM5}).

\subsection{Time integration}

One of the strength of pseudo-spectral methods is that they replace a set of partial differential equations (PDE) by a larger set of ordinary differential equations (ODE) for the unknown collocation points or spectral coefficients. Schematically, it can be written  as
\begin{equation}
  \label{eq:ODE}
  \frac{d\bmath{u}}{dt} = f(t,\bmath{u})
\end{equation}
with appropriate initial and boundary conditions. $\bmath{u}$ represents the vector of unknown functions either evaluated at the collocation points or the spectral coefficients. We use a fourth-order Runge-Kutta scheme advancing the unknown functions~$\bmath{u}$ in time. See also the discussion in \cite{Hesthaven2008} for more details about other time integration schemes especially those called strong stability preserving Runge-Kutta methods including the popular schemes of order two and three (SSPRK2,3).

\subsection{Initial conditions}

The rotation of the neutron star is switched on smoothly as in \cite{2012MNRAS.424..605P}. Its spin frequency increase slowly from zero in order to avoid the formation of sharp gradients. This would be especially true at time $t=0$ where there is no electric field outside the star but right on its surface. Taking an evolution of the spin frequency as
\begin{equation}
 \Omega(t) = 
\begin{cases}
\sin^2\left(\frac{t}{8}\right) \text{ for } t\leq4\,\upi \\ 
1  \text{ for } t\geq4\,\upi
\end{cases}
\end{equation}
therefore starting at a null value avoids the initial discontinuity in the electric field. The spin frequency as well as its first derivative are smooth at the initial time of the simulation~$t=0$. No gradient or sharp features are expected. We next switch to a discussion of the results.

\section{NON RELATIVISTIC TESTS}
\label{sec:Tests}

For the remaining of the paper, we adopt the following normalization: the magnetic moment of the star is equal to unity, therefore $B\,R^2=1$, as well as the stellar angular velocity and the speed of light, $\Omega = c = \varepsilon_0 = \mu_0 = 1$, therefore the light-cylinder radius is $\rlight=1$.

We start with a discussion about the non-relativistic monopole solutions in vacuum but also in the force-free limit. Interestingly analytical closed expressions do exist in these cases. They are very valuable solutions to check the correctness and accuracy of our code. The general-relativistic rotator will be treated in section~\ref{sec:Results}.

Although a discontinuous Galerkin method is intended to do better than second order in space, in this paper we only show results with $N_p=1$ i.e. use linear polynomial interpolation of the unknown fields. Indeed, so far we only implemented TVD and TVBM limiters which fall down to first order at shocks or when a limiting procedure is applied. We plan to add higher order slope limiters in the near future such as the moment limiter described in \cite{Biswas:1994:PAF:179015.179030}. Fortunately we already get accurate solution with linear polynomials.

\subsection{Vacuum monopole solution}

We tested our code against some well known analytical solutions. The starting point is the vacuum monopole field for which the Poynting flux is equal to zero. The solution has been presented in section~\ref{sec:Monopole}. The analytical solution is exact and easy to compare with the output of our simulations.

We start our computation with a non rotating monopole magnetic field, $\Omega=0$, and zero electric field outside the star, except for the crust where we enforce the inner boundary condition, see equation~(\ref{eq:CLimites}). Note however that due to our special profile of $\Omega(t)$, the electric field at the surface of the star is initially equal to zero. It will slowly increase to its maximal value reached at a normalized time $t=4\,\upi$.

We performed simulations with different spin frequencies of the neutron star corresponding to several ratio between stellar radius~$R$ and light cylinder radius~$\rlight$ such that $\rlight/R = \{2,10\}$ and between the artificial outer boundary and the light-cylinder $R_{\rm out}/\rlight = \{10,100,1000\}$. Obviously the resolution of the grid should be highest for the largest domain in radius with $r/\rlight\in[0.1,1000]$. A minimum resolution of $K \times N_{\rm p} \times N_\vartheta = 128\times1\times4$ was necessary. Actually, throughout the paper, we will show results with a higher resolution of $K \times N_{\rm p} \times N_\vartheta = 256\times1\times8$.  Because of the axisymmetry of the problem, a Fourier transform in the azimuthal direction is not necessary, so we simply put $N_\varphi=1$. We let the system evolve until it reaches a stationary state inside the simulation box. Thus the final time strongly depends on the location of the outer boundary, it can be as high as $t_{\rm final} = 300\,\upi$ for $R_{\rm out}=1000\,\rlight$. 

\begin{figure*}
  \centering
  \begin{tabular}{cc}
  \includegraphics[width=0.5\textwidth]{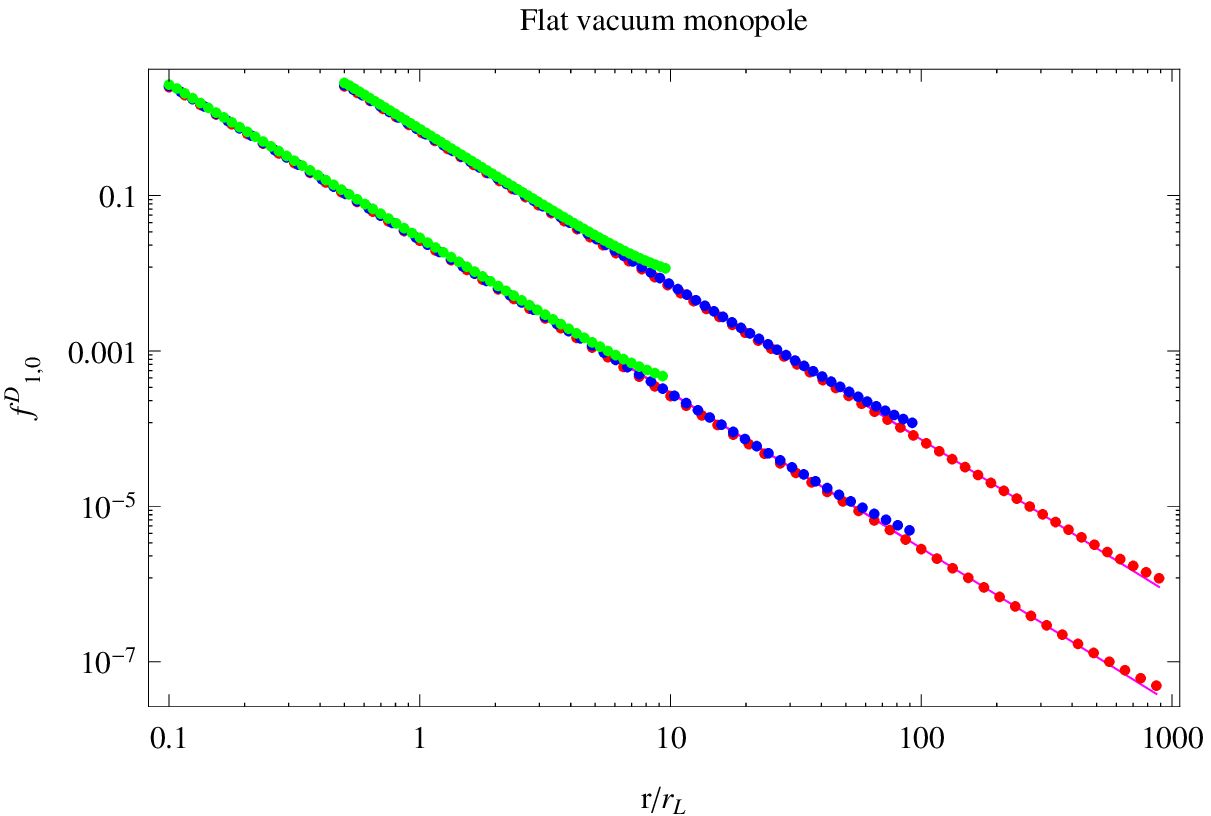} &
  \includegraphics[width=0.5\textwidth]{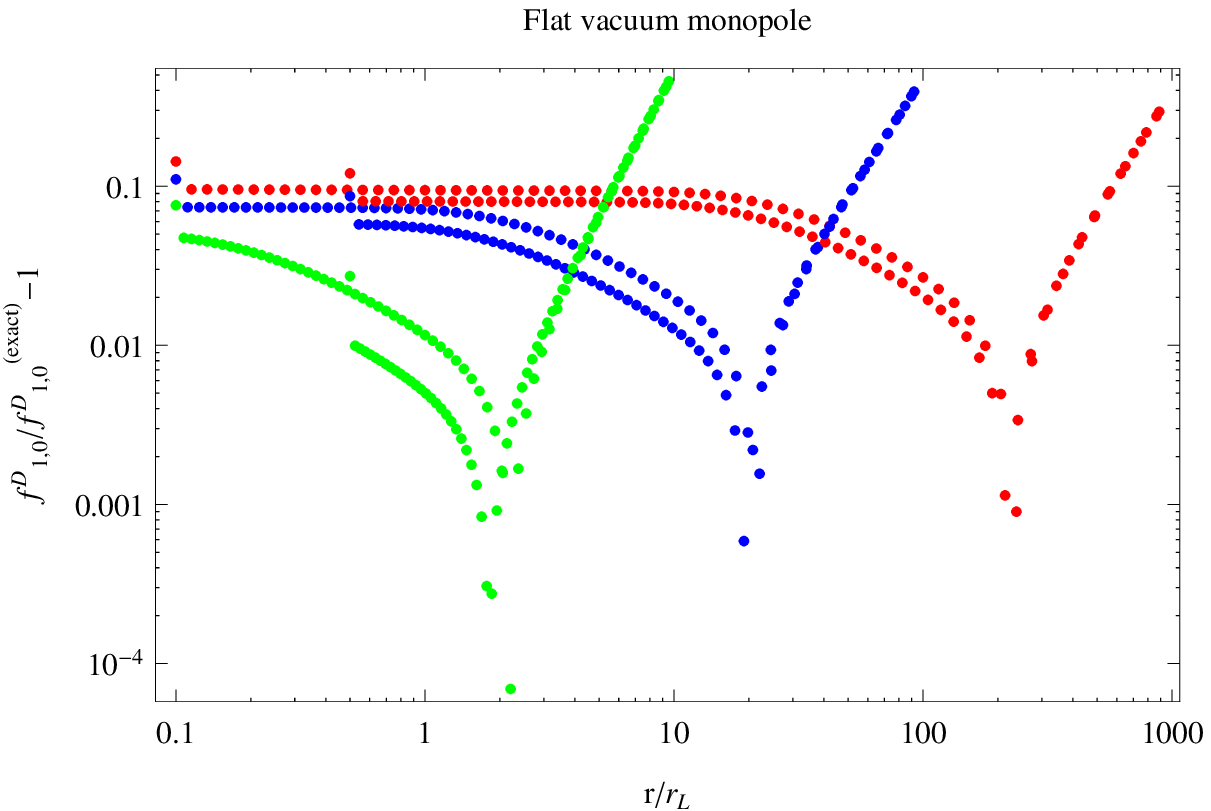}
  \end{tabular}
  \caption{The function~$f^D_{1,0}(r)$ of the vacuum monopole solution for $R_{\rm out}/\rlight=\{10,100,1000\}$ and a ratio $\rlight/R=\{2,10\}$. The time-dependent simulation in red, green and blue dots is compared to the exact analytical solution in solid red lines. They are hardly distinguishable as can be checked from the relative error on the right panel.}
  \label{fig:Monopole_j0_fD10_flat}
\end{figure*}

\begin{figure}
  \centering
  \includegraphics[width=0.5\textwidth]{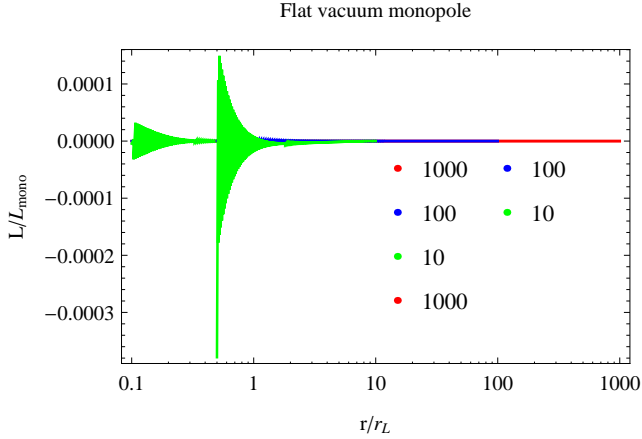}
  \caption{The Poynting flux for the vacuum monopole solution for $R_{\rm out}/\rlight=\{10,100,1000\}$ and a ratio $\rlight/R=\{2,10\}$. As expected it is zero within numerical accuracy.}
  \label{fig:Monopole_j0_poynting_flat}
\end{figure}

In the non-relativistic monopole solution, the magnetic field remains unchanged. The only relevant quantity to check is the coefficient~$f^D_{1,0}(r)$ for the electric field. It is understood that all other coefficients should be equal to zero. In figure~\ref{fig:Monopole_j0_fD10_flat} we show this coefficient~$f^D_{1,0}(r)$  on the left panel and its relative error on the right panel for several sets of parameters. Note that it is plotted on a log-log scale in order to make more visible the outer part of the function. A careful investigation of this outer part shows a slight deviation of the computed solution with respect to the analytical solution. Let us assume that the solution is accurate if the relative error is less than the one reached close to the neutron star surface. Then if $R_{\rm out}=10\,\rlight$ the computed solution becomes inaccurate above $\approx5\,\rlight$ but if $R_{\rm out}=100\,\rlight$ then the discrepancy starts at $\approx50\,\rlight$ and finally for $R_{\rm out}=1000\,\rlight$ the inaccuracy starts at $\approx500\,\rlight$. This behaviour clearly indicates an influence of the location of the outer boundary on the numerical solution. Such artifact can only be removed by moving away the artificial outer boundary.  Using a smaller time step will not help to improve the accuracy or to remove the outer boundary influence. Indeed, we run the same simulations with a time step 2.5 or 5 times smaller than the one  presented here for relative error. We have not noticed any changes in this error so these plots are not shown to avoid congesting the figures.
The corresponding Poynting flux is shown in figure~\ref{fig:Monopole_j0_poynting_flat}. As expected it is very close to zero as it should be. The accuracy is better than $10^{-3}$ in the whole simulation box whatever its size. Note that even if the solution is inaccurate at large distances, the associated Poynting flux, although having large errors, remains close to zero. This is explained by the fact that the electromagnetic field in those region is weak. It is impossible to compute the relative error in the Poynting flux because the exact value should be zero.

\begin{figure*}
  \centering
\begin{tabular}{cc}
  \includegraphics[width=0.5\textwidth]{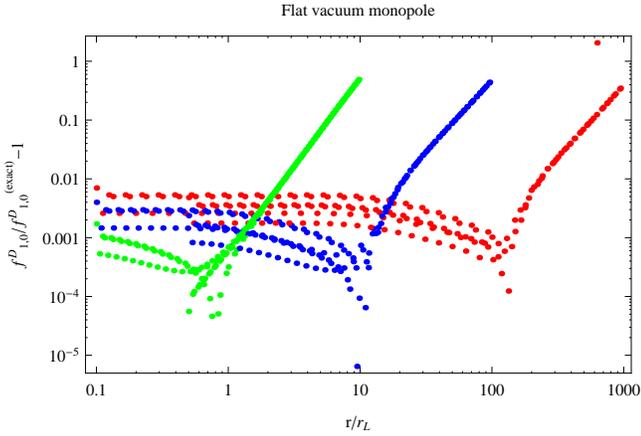} &
  \includegraphics[width=0.5\textwidth]{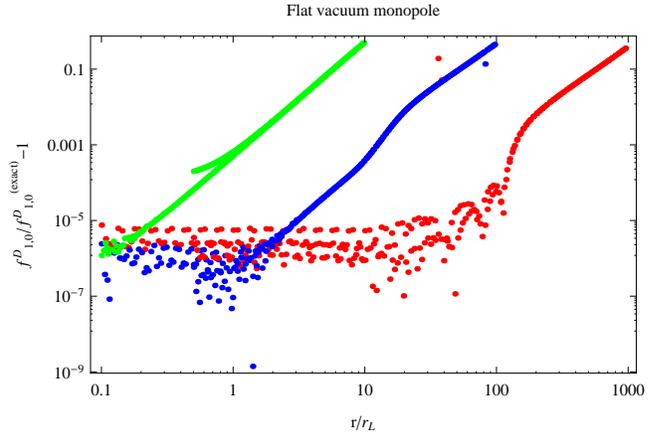}
\end{tabular}
  \caption{The relative error of the function~$f^D_{1,0}(r)$ for the vacuum monopole with a second order polynomial approximation, $N_p=2$ on the left panel, and a fourth order polynomial approximation, $N_p=4$ on the right panel. These have to be compared with the linear approximation in fig.~\ref{fig:Monopole_j0_fD10_flat}.}
  \label{fig:Monopole_j0_fD10_err_flat}
\end{figure*}

To conclude with the vacuum case, note that Maxwell equations become linear. Therefore we do not need to apply a strong limiting in the radial direction. In that case, we can use higher order spatial expansions of the unknown fields without destroying the high order of the method. This has been done for instance with a quadratic $N_p=2$ and a fourth order $N_p=4$ polynomial expansion. Results of such simulations are shown in figure~\ref{fig:Monopole_j0_fD10_err_flat} where the relative error in the function $f^D_{1,0}$ is plotted and has to be compared with the corresponding plot in fig.~\ref{fig:Monopole_j0_fD10_flat} with $N_p=1$. We used the same number of cells in each computation. It is clear that higher order methods are much more accurate. This demonstrates the need for limiters that do preserve the high order accuracy of discontinuous Galerkin schemes.

The above results demonstrate that the code is able to catch accurate solutions of the vacuum electromagnetic field with appropriate boundary conditions on the perfectly conducting star and at large distances. As we now discuss, in the force-free limit the code also gives accurate solutions.

\subsection{Force-free monopole solution}

\begin{figure*}
  \centering
  \begin{tabular}{cc}
  \includegraphics[width=0.5\textwidth]{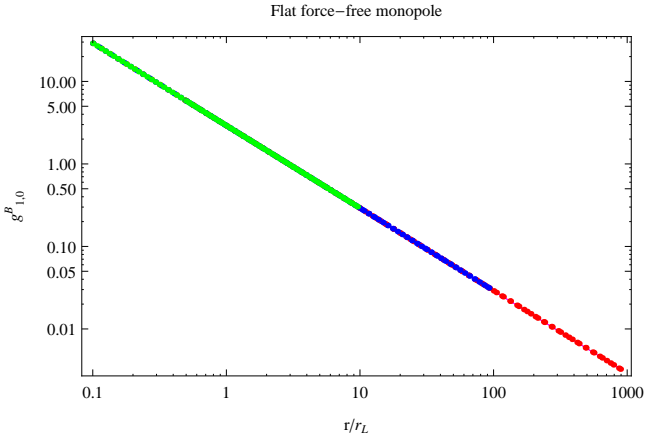} &
  \includegraphics[width=0.5\textwidth]{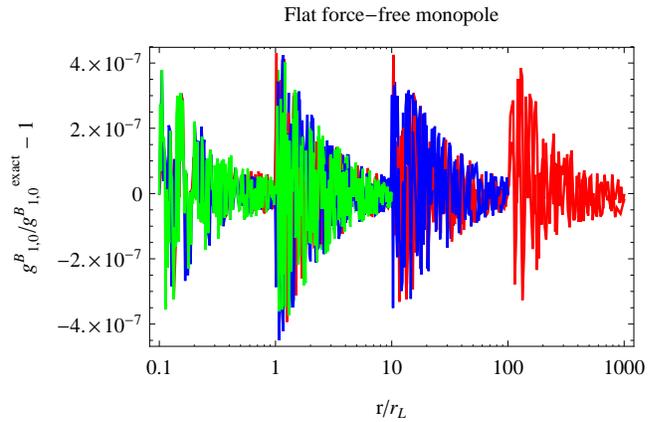}
  \end{tabular}
  \caption{Magnetic field coefficient $g^B_{1,0}$ for the force-free monopole solution for  $R_{\rm out}/\rlight=\{10,100,1000\}$ and a ratio $\rlight/R=\{2,10\}$, left panel. All the curves overlap and are graphically undistinguishable. $g^B_{1,0}$ is compared to the exact analytical expression through the relative error $g^B_{1,0}/g^{B(exact)}_{1,0}-1$, right panel.}
  \label{fig:Monopole_j1_gB10_flat}
\end{figure*}

\begin{figure}
  \centering
  \includegraphics[width=0.5\textwidth]{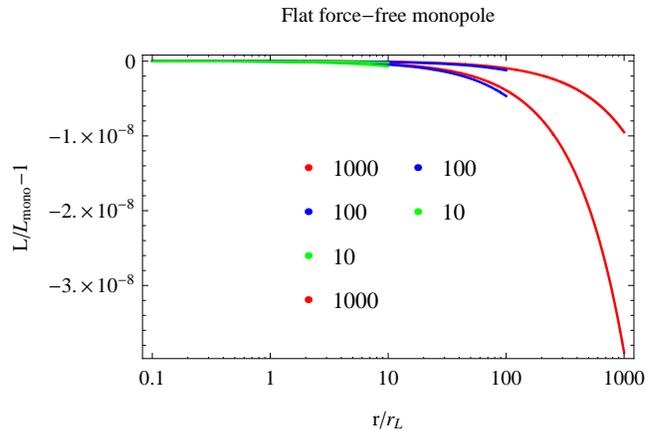}
  \caption{The Poynting flux relative error for the force-free monopole solution for $R_{\rm out}/\rlight=\{10,100,1000\}$ and a ratio $\rlight/R=\{2,10\}$. As expected it is equal to $L_{\rm mono}$ to very high precision, better than 7~digits.}
  \label{fig:Monopole_j1_poynting_flat}
\end{figure}

Next we tackle the problem of an axisymmetric force-free flow known as the monopole field introduced by \cite{1973ApJ...180L.133M}.  We recall that this monopole solution is given by
\begin{equation}
  \bmath{B} = B_{\rm L} \, \frac{\rlight^2}{r^2} \, \er - B_{\rm L} \, \frac{\rlight}{r} \, \sin\vartheta \, \ephi
\end{equation}
In terms of a vector spherical harmonic (VSH) expansion, this magnetic field is expressed as
\begin{equation}
 \bmath{B} = B_{\rm L} \, \frac{\rlight^2}{r^2} \, \er + g_{1,0}^{B(exact)}(r) \, \mathbf{\Phi}_{10}  
\end{equation}
where
\begin{equation}
  \label{eq:gB10}
  g_{1,0}^{B(exact)}(r) = \sqrt{\frac{8\pi}{3}} \, B_L \, \frac{\rlight}{r}
\end{equation}
all other coefficients being equal to zero. The associated Poynting flux is
\begin{equation}
  \label{eq:L_mono}
  L_{\rm mono} = \frac{8\,\pi}{3\,\mu_0\,c^3} \, \Omega^4 \, B_{\rm L}^2 \, \rlight^6
\end{equation}
The initial set up is the same as in the previous paragraph. We only add a source term represented by the force-free current given by equation~(\ref{eq:CourantFFE}). During the evolution of the electromagnetic field, it is easy to show that the component $B_r$ remains constant in time and that only the $B_\varphi$ component is present with the coefficient $g_{1,0}^{B(exact)}(r)$. The numerical value of this coefficient is shown in the left panel of figure~\ref{fig:Monopole_j1_gB10_flat}. Moreover, in order to prove the accuracy of our code, we plot the ratio $g^B_{1,0}/g_{1,0}^{B(exact)}(r)$ and compare it to unity as depicted in figure~\ref{fig:Monopole_j1_gB10_flat}, right panel. The accuracy is better than 6 digits in the whole computational domain.
For completeness we also plot the Poynting flux obtained from the simulations as shown in figure~\ref{fig:Monopole_j1_poynting_flat}. From the analytical solution, we known that the Poynting flux is a constant, irrespective of the size of the neutron star. This is indeed what we found. In normalized units, the Poynting flux is equal to unity whatever the ratio $\rlight/R$ and whatever the location of the outer boundary. The result is very accurate, better than 7~significant digits. Interestingly, contrary to the vacuum monopole field, the force-free solution does not suffer from the location of the outer boundary. We always found the exact analytical expression (to high numerical accuracy) in the whole simulation box. Thus a small digression about these outer boundaries is in order as exposed in the next paragraph.

\subsection{Influence of the location of the outer boundary}

Imposing exact outgoing wave boundary conditions on a sphere of finite radius is a tedious work. Indeed \cite{2004JCoPh.197..186N} showed that the Sommerfeld radiation condition is only valid for the monopole field. For dipolar or even multipolar structures, restricting the infinite domain to a sphere of radius~$R_{\rm out}$ will lead to some deviation from a perfect outgoing wave. To elucidate the influence of the location of this outer sphere, we looked at the error of the Poynting flux with respect to the location of $R_{\rm out}$ defined by
\begin{equation}
\label{eq:erreur_dout} 
\epsilon = \left|\frac{L_{\rm ana} - L_{\rm num}}{L_{\rm ana}}\right|
\end{equation}
where $L_{\rm ana}$ and $L_{\rm num}$ are the analytical and numerical Poynting fluxes respectively. We report our results in this brief paragraph for the flat space-time, choosing a radius of the neutron star equal to $\rlight/R=2$ and $R_{\rm out}/\rlight = \{10,100,100\}$. For the vacuum or force-free field we know exact solutions. As we already showed in figure~\ref{fig:Monopole_j0_fD10_flat} there is a slight influence for the vacuum field. Nevertheless we did not found any influence on the force-free solution. 

We demonstrated in this section that our pseudo-spectral discontinuous Galerkin code is mature and able to compute accurately vacuum as well as force-free electromagnetic fields in flat space-time. Boundary conditions have been implemented in an efficient way avoiding spurious reflections and artificial inner boundaries as usually required for finite difference/volume methods. Before looking at the general-relativistic solution we finish the test in flat space-time by a discussion about the important situation where a current sheet is present in the solution.

\subsection{Split monopole solution}

Our first intention to implement the discontinuous Galerkin method was to handle multi-domain computational boxes, allowing for non-uniform grids and therefore larger scales. However, this method is also well suited for the study of solutions presenting discontinuities. So we decided to test our code against a magnetic field structure showing a current sheet in the equatorial plane as for instance in the split monopole field. It is well known that analytically the solution is made of two half monopole fields of opposite ``magnetic charge'' separating the space into two hemispheres where the above force-free monopole applies separately. We have not met any particular problem to deal with this discontinuous solution. Let us investigate in more details the split monopole.

At the surface of the star, the radial component of the magnetic field reverses polarity at the equator. It therefore represents a step function in the $\vartheta$ variable on which we perform a series expansion. This jump will introduce the well-known Gibbs phenomenon and decrease the convergence rate to the worst case: first order. The Gibbs phenomenon produces an associated overshoot in $B_r$ that do not decrease by increasing the number of terms in the expansion, i.e. $N_\vartheta$. This is proved rigorously mathematically. The filtering explained in the code description section will help to enforce a lowering of these spurious oscillations. In any case, the current sheet does not pollute or even destroy the solution in the simulation domain.

\begin{figure}
  \centering
  \includegraphics[width=0.5\textwidth]{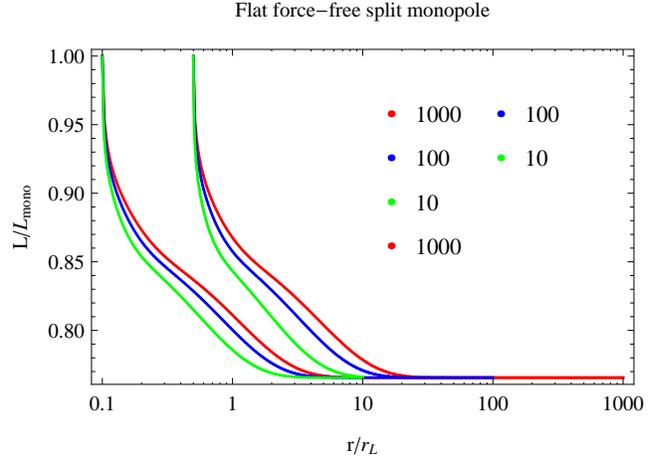}
  \caption{Normalized Poynting flux~$L/L_{\rm mono}$ across the sphere of radius~$r$ where $L$ is evaluated from eq.~(\ref{eq:Poynting}) and $L_{\rm mono}$ given by eq.~(\ref{eq:L_mono}). Dissipation reaches up to 25\%. The inset legend corresponds to the ratio $R_{\rm out}/\rlight = \{10,100,1000\}$ and $\rlight/R=\{2,10\}$. Note the logarithm scale in radius.}
  \label{fig:split_j1_poynting_flat}
\end{figure}

\begin{figure*}
  \centering
  \begin{tabular}{cc}
  \includegraphics[width=0.5\textwidth]{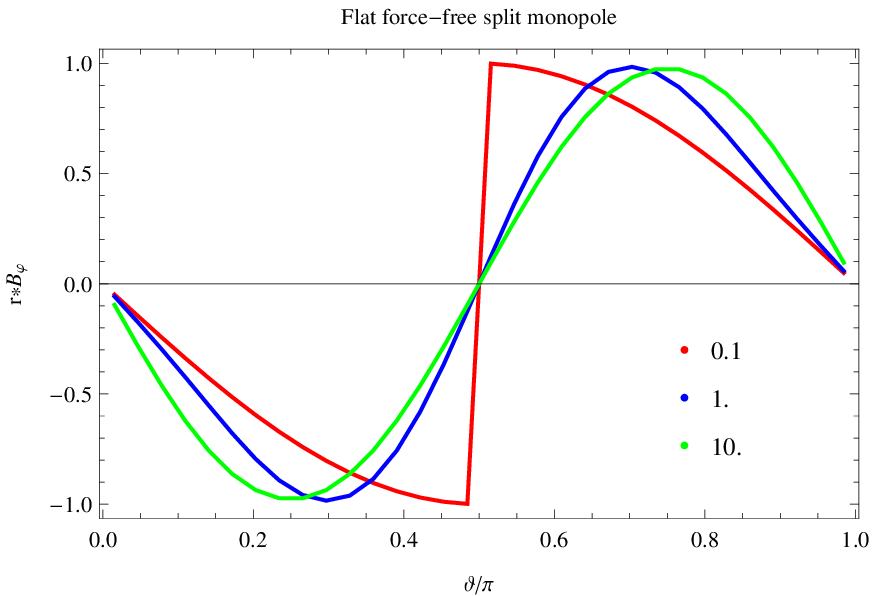} &
  \includegraphics[width=0.5\textwidth]{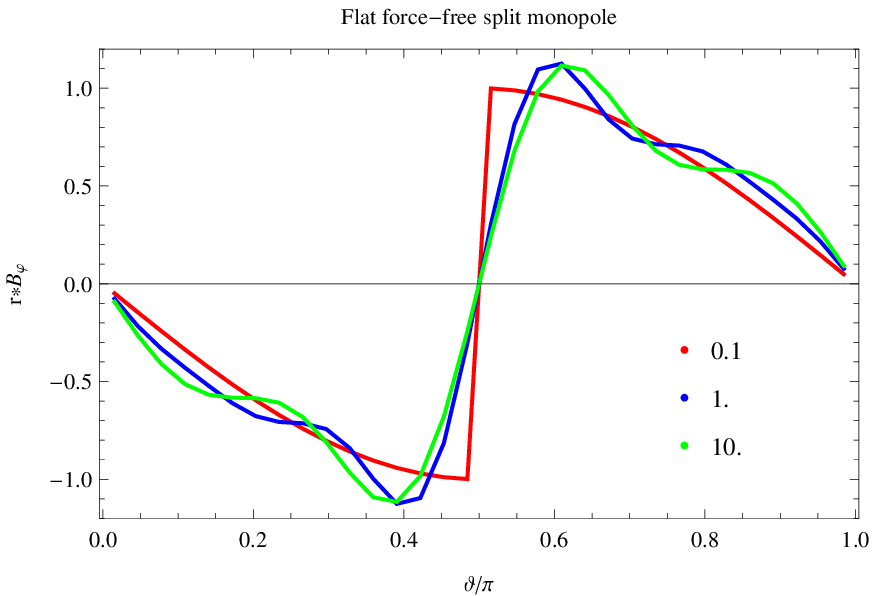}
  \end{tabular}
  \caption{Azimuthal component of the magnetic field~$B_\varphi$ for the split monopole solution at three different radii, at the neutron star surface $r=R$ (red curve), at some point inside the simulation box $r=\rlight$ (blue curve) and at the outer boundary $r=10\,\rlight$ (green curve), using different filtering orders, to the left, $\beta=4$ and to the right $\beta=8$. The parameters are $N_\vartheta=32$, $R_{\rm out}/\rlight=10$ and $\rlight/R=10$. $B_\varphi$ is multiplied by $r$ to get ride of the radial dependence. In the exact analytical solution, all three curves should overlap.}
  \label{fig:split_j1_B_p_flat}
\end{figure*}

The Poynting flux is shown in fig.~\ref{fig:split_j1_poynting_flat} for the ratio $R_{\rm out}/\rlight = \{10,100,1000\}$ and $\rlight/R=\{2,10\}$. Theoretically, we know that this flux should be equal to the force-free monopole luminosity, so in normalized units it should equal to unity. But the filtering and limiting procedures, useful to prevent strong numerical oscillations and possible non-linear instabilities, introduce some nonphysical dissipation. This is clearly recognized in fig.~\ref{fig:split_j1_poynting_flat} where the computed Poynting flux decreases with radius. The rate of dissipation can be controlled by the resolution of the simulation and the filtering. This is shown in fig.~\ref{fig:split_j1_B_p_flat} where the azimuthal component~$B_\varphi$ is plotted against the colatitude~$\vartheta$ at three different radii, namely at the neutron star surface, at some point inside the simulation box and at the outer boundary. We recognize the Gibbs phenomenon through its oscillatory nature in the vicinity of the discontinuity. The solution becomes more accurate when we increase the number of coefficients in the $\vartheta$ expansion and/or if we reduce the influence of the filtering on the lowest multipole coefficients.

The dissipation outside the light-cylinder is close to 25\%. We plan to reduce this strong dissipation by replacing the fist order TVBM limiter by higher order filtering and increasing the number of discretization points in both directions. Nevertheless, this improvement of our code is left for future work.

\section{GENERAL-RELATIVISTIC MONOPOLE SOLUTIONS}
\label{sec:Results}

We now present new results about the monopole force-free solution in general relativity. We adopt the fixed background metric for a slowly rotating neutron star in Boyer-Lindquist coordinates as described in section~\ref{sec:Modele}.

The same spin frequencies than those for the non-relativistic solutions are used, corresponding to $R_{\rm out}/\rlight=\{10,100,1000\}$ whereas the spin frequency is such that $\rlight/R=\{2,10\}$. The compactness, typical of a neutron star, is set to~$\Xi^{-1}=R/\Rs=2$.

\subsection{Vacuum monopole}

\begin{figure}
  \centering
  \includegraphics[width=0.5\textwidth]{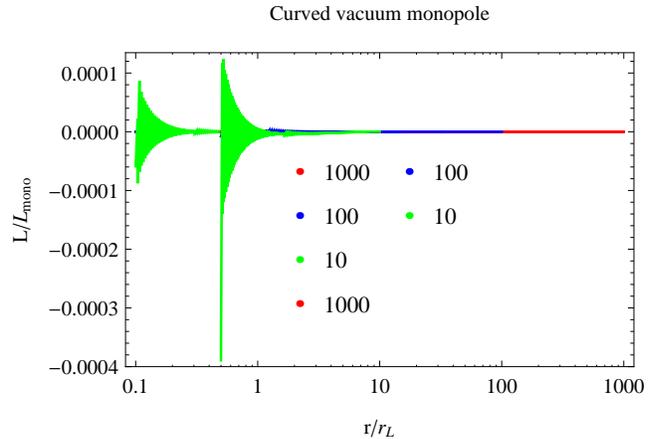}
  \caption{Normalized Poynting flux~$L/L_{\rm mono}$ across the sphere of radius~$r$ where $L$ is evaluated according to equation~(\ref{eq:Poynting}) and  $L_{\rm mono}$ is given by equation~(\ref{eq:L_mono}). The computed flux vanishes as expected, within the numerical precision of the algorithm. The solution settled down to a stationary state. The inset legend corresponds to the ratio $R_{\rm out}/\rlight = \{10,100,1000\}$. Note the logarithm scale in radius.}
  \label{fig:monopole_j0_poynting_curved}
\end{figure}

Approximate expressions for the vacuum monopole field in general relativity are given as outlined in section~\ref{sec:Monopole}. No outgoing electromagnetic wave propagating into vacuum space exists except for a transient regime relaxing to the stationary state. The Poynting flux as seen by an observer at infinity therefore vanishes. We checked this assertion by plotting the Poynting flux in figure~\ref{fig:monopole_j0_poynting_curved} according to equation~(\ref{eq:Poynting}). Different runs are shown corresponding to increasing size of the simulation box, namely for the set of ratio $\rlight/R = \{10,100,1000\}$. The Poynting flux vanishes everywhere to very good accuracy. Moreover the electromagnetic field evolved to a steady state without reflection at the outer boundary. Our characteristics compatibility method used in flat space-time does also give good results in a curved space-time, when the outer boundary is kept far from the light cylinder, justifying its numerical use. Note however that the outer edge of the box is not rigorously transparent to electromagnetic waves as was already the case with the flat vacuum monopole. The only remedy to this inaccuracy is to enlarge the box size at the expense of computational time due to the propagation delay between the star and the outer boundary and due to the requirement of higher grid resolutions.

\begin{figure*}
  \centering
\begin{tabular}{cc}
  \includegraphics[width=0.5\textwidth]{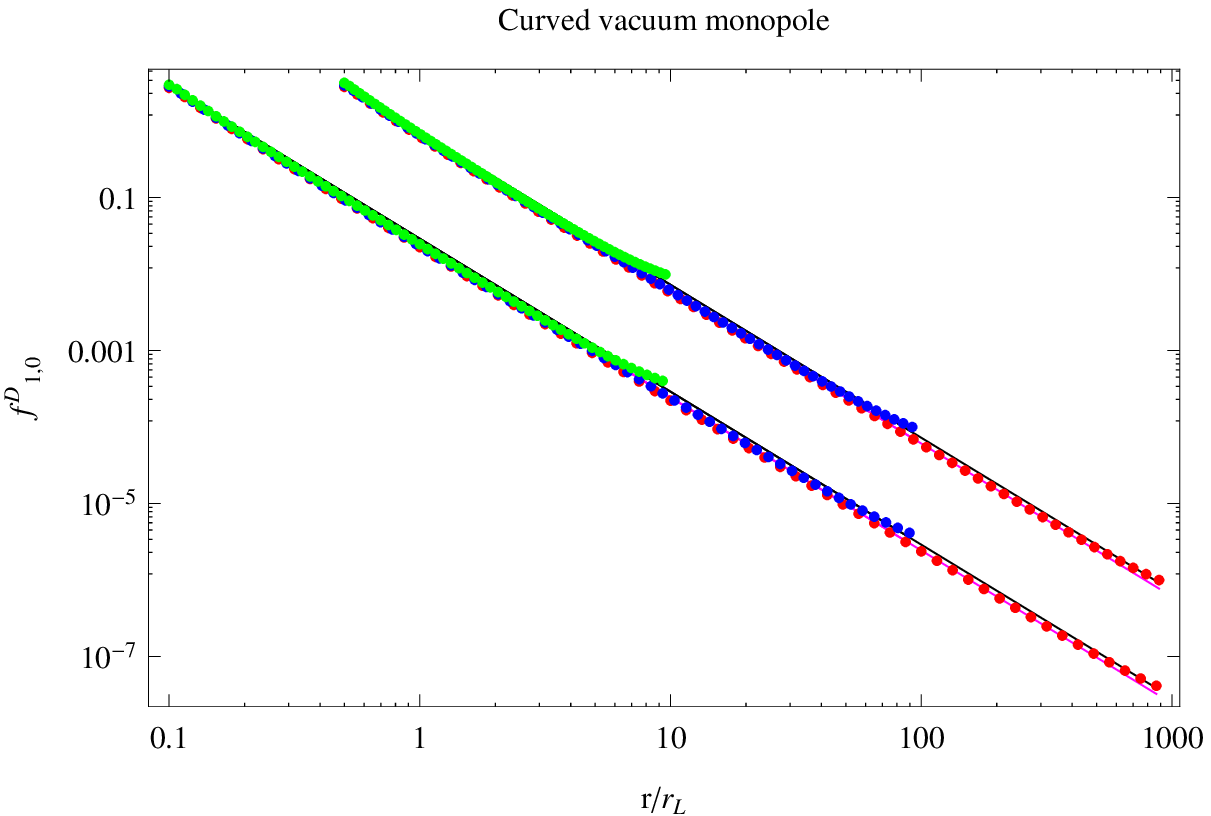} &
  \includegraphics[width=0.5\textwidth]{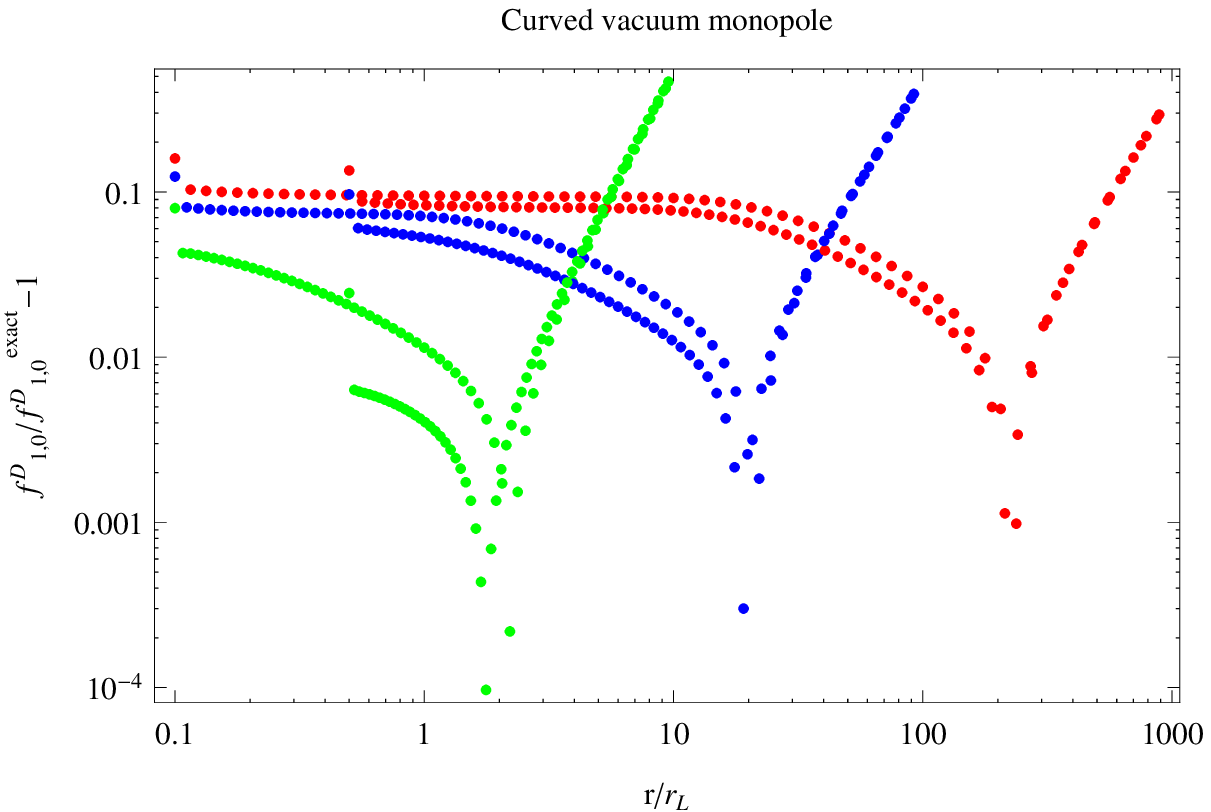}
\end{tabular}
  \caption{The function~$f^D_{1,0}(r)$ for the vacuum monopole solution in general relativity, on the left panel, and it relative error on the right panel, with $R_{\rm out}/\rlight=\{10,100,1000\}$ and a ratio $\rlight/R=\{2,10\}$. In the left panel, the time-dependent simulations in red, green and blue dots are compared to the first order analytical solution in solid magenta lines. They are hardly distinguishable. For completeness the flat vacuum solution is shown in black solid lines.}
  \label{fig:monopole_j0_fD10_curved}
\end{figure*}

In order to give an estimate of the accuracy of our computed solution, the first order approximation of $f^D_{1,0}$ given by the analytical expression equation~(\ref{eq:fD10}) is compared with the output of the pseudo-spectral discontinuous Galerkin code. Results are shown in figure~\ref{fig:monopole_j0_fD10_curved} for the function $f^D_{1,0}$ itself, on the left panel, and its relative error on the right panel. We find good agreement between both functions. Although the time-dependent simulations contain multipolar electromagnetic fields with $l>1$, the computed solution do not differ much from the analytical expression containing only the dipolar electric field $l=1$. As expected, the corrections induced by the mulitpolar components remain negligible. General-relativistic effects stay on a low level. The flat vacuum function almost overlaps the curved space-time counterpart.

\begin{figure*}
  \centering
\begin{tabular}{cc}
  \includegraphics[width=0.5\textwidth]{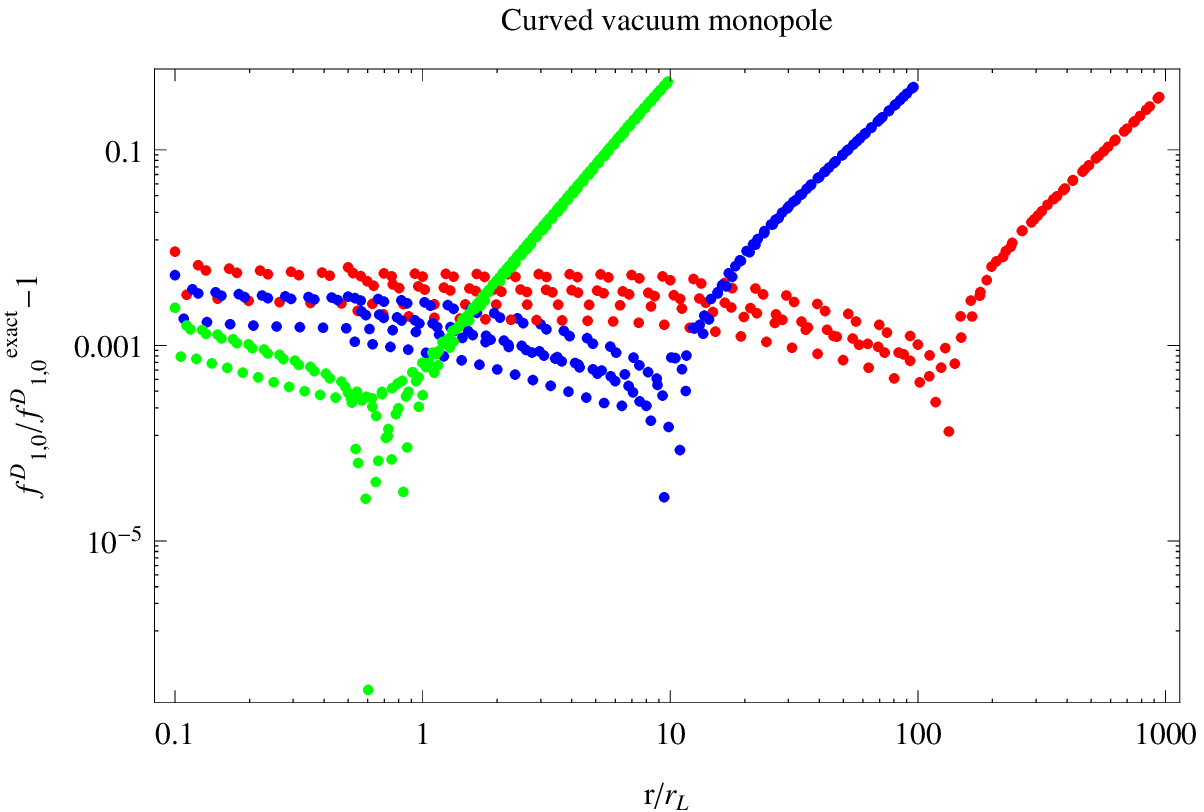} &
  \includegraphics[width=0.5\textwidth]{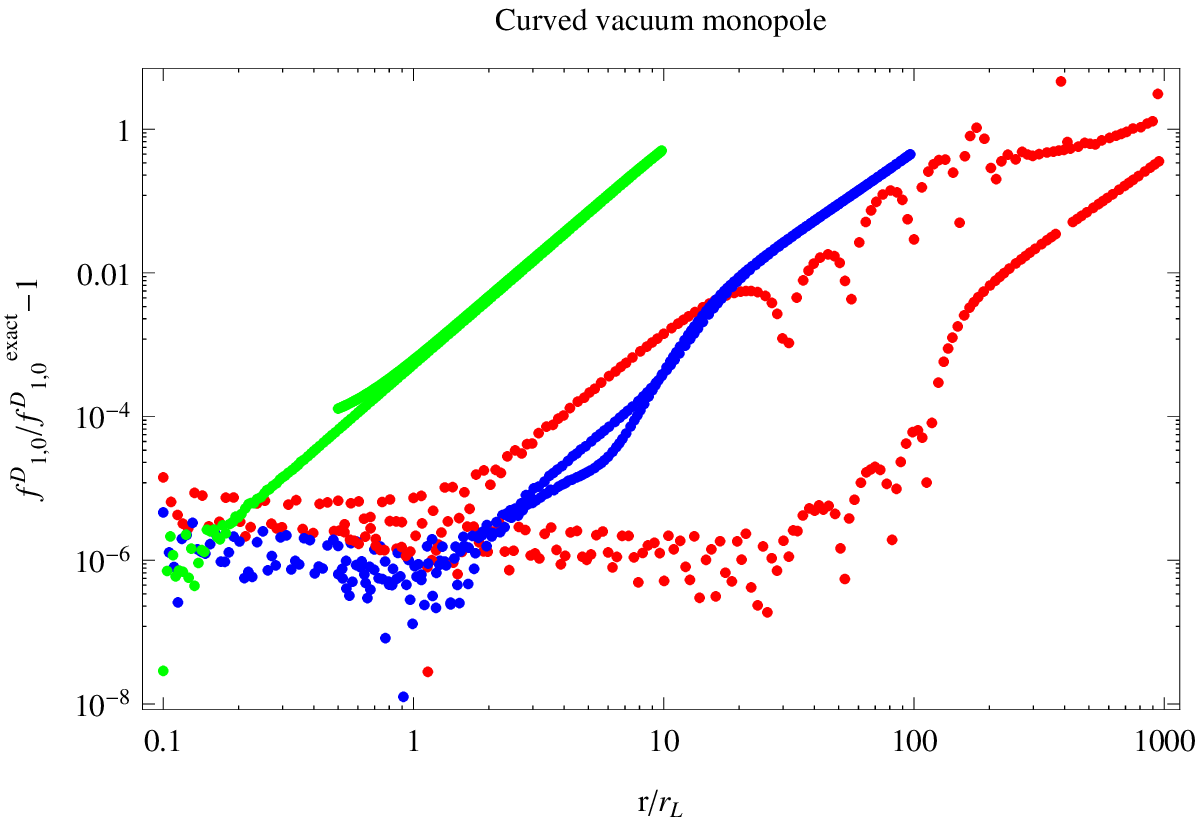}
\end{tabular}
  \caption{The relative error of the function~$f^D_{1,0}(r)$ for the vacuum monopole with a second order polynomial approximation, $N_p=2$ on the left panel, and a fourth order polynomial approximation, $N_p=4$ on the right panel. These have to be compared with the linear approximation in fig.~\ref{fig:monopole_j0_fD10_curved}.}
  \label{fig:monopole_j0_fD10_err_curved}
\end{figure*}

In the general-relativistic case too, a higher order spatial expansion remains more accurate than a low order one. To demonstrate it, we performed here again simulations with $N_p=2$ or $N_p=4$. Results are shown in figure~\ref{fig:monopole_j0_fD10_err_curved} and should be compared to the linear approximation in fig.~\ref{fig:monopole_j0_fD10_curved} with $N_p=1$.

To sum up, we proved that our code is able to computed accurate solutions of the electromagnetic field in a fixed curved geometry. We eventually switch to the most interesting case, the general-relativistic force-free monopole field.

\subsection{Force-free monopole}

\begin{figure}
  \centering
  \includegraphics[width=0.5\textwidth]{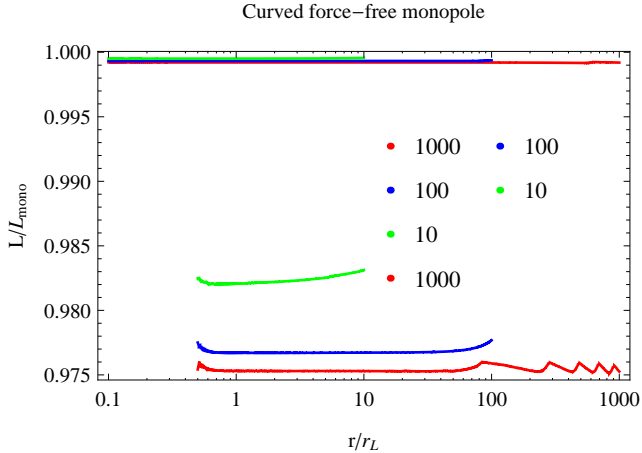}
  \caption{Normalized Poynting flux~$L/L_{\rm mono}$ across the sphere of radius~$r$ where $L$ is evaluated from eq.~(\ref{eq:Poynting}) and $L_{\rm mono}$ given by eq.~(\ref{eq:L_mono}). The computed flux is constant as expected. The solution settled down to a stationary state. The inset legend corresponds to the ratio $R_{\rm out}/\rlight = \{10,100,1000\}$ and $\rlight/R=\{2,10\}$. Note the logarithm scale in radius.}
  \label{fig:monopole_j1_poynting_curved}
\end{figure}

\begin{figure}
  \centering
  \includegraphics[width=0.5\textwidth]{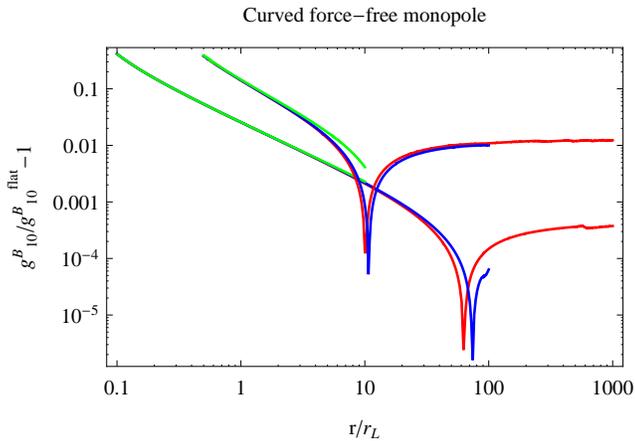}
  \caption{Deviation of the general-relativistic magnetic field coefficient $g^B_{1,0}$ from its Newtonian version in the force-free monopole with $R_{\rm out}/\rlight=\{10,100,1000\}$ and a ratio $\rlight/R=\{2,10\}$. It is compared to the flat monopole through the relative error $g^B_{1,0}/g^{B(flat)}_{1,0}-1$.}
  \label{fig:monopole_j1_gB10_curved}
\end{figure}

Current wisdom assumes that pulsars are neutron stars surrounded by relativistic plasmas of electron/positron pairs. If the pulsed radio emission comes from the magnetic poles, then we should look for accurate configurations of the magnetic field in the vicinity of the neutron star including curved space-time and plasma screening effects. This last paragraph is intended to bring us one step closer to this difficult task. As a starting point, we envisage a monopolar magnetic field instead of the more traditional dipolar structure. A detailed investigation of the general-relativistic dipole force-free magnetosphere is left for upcoming work. A simple prescription including the plasma current is based on the force-free approximation as explained in section~\ref{sec:Modele}. The simulation set up is the same as in vacuum except that the force-free current is switched on. We summarize the results by showing the Poynting flux for the different runs as presented in figure~\ref{fig:monopole_j1_poynting_curved}. We found that the power radiated does not significantly deviate from its Minkowski version. For the case $\rlight/R=10$, the normalized Poynting flux is equal to unity with 0.1\%. It is the same as the Michel monopole solution given in \cite{1973ApJ...180L.133M}. However, for the case $\rlight/R=2$, we observe a deviation from the flat space-time monopole Poynting flux around 2\%. This decrease of the luminosity is a direct consequence of the frame dragging effect, being more pronounced in that case. If we artificially switch off the frame dragging effect by setting $\bmath \beta=0$, we would retrieve to good accuracy the Newtonian case, independently of the ratio $\rlight/R$.
For completeness the coefficient $g^B_{1,0}$ is also compared to its flat space-time version in figure~\ref{fig:monopole_j1_gB10_curved}. General relativity distorts the field sensitively close to the neutron star.
As expected, general-relativistic effects are important only close to the neutron star surface where curvature and frame-dragging are significant. The lowest order azimuthal magnetic field geometry is distorted with respect to its flat counterpart. Nevertheless the Poynting flux as measured by a distant observer is not significantly affected by such perturbations. Even if multipolar components are present, they remain at a low level compared to the dominant multipole electric and magnetic field.

\subsection{Split monopole}

\begin{figure}
  \centering
  \includegraphics[width=0.5\textwidth]{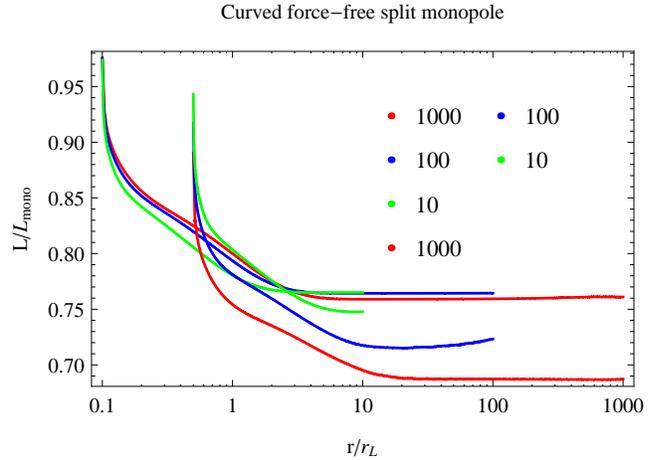}
  \caption{Normalized Poynting flux~$L/L_{\rm mono}$ across the sphere of radius~$r$ where $L$ is evaluated from eq.~(\ref{eq:Poynting}) and $L_{\rm mono}$ given by eq.~(\ref{eq:L_mono}). The computed flux is dissipated up to 30\%. The inset legend corresponds to the ratio $R_{\rm out}/\rlight = \{10,100,1000\}$ and $\rlight/R=\{2,10\}$. Note the logarithm scale in radius.}
  \label{fig:split_j1_poynting_curved}
\end{figure}

The study of the split monopole in flat space-time can be repeated in general relativity. A typical set of runs is shown in fig.~\ref{fig:split_j1_poynting_curved}. The same remarks as for the Newtonian split monopole hold here. Dissipation is again introduced in order to minimize the effect of the Gibbs phenomenon. At the stellar surface the Poynting flux is maximal and close to the true value but as soon as we depart from the stellar surface, energy is dissipated and diminishes the measured outgoing Poynting flux. Energy is dissipated up to 30\%. Nevertheless the solution settled down to a stationary state.

\begin{figure*}
  \centering
  \begin{tabular}{cc}
  \includegraphics[width=0.5\textwidth]{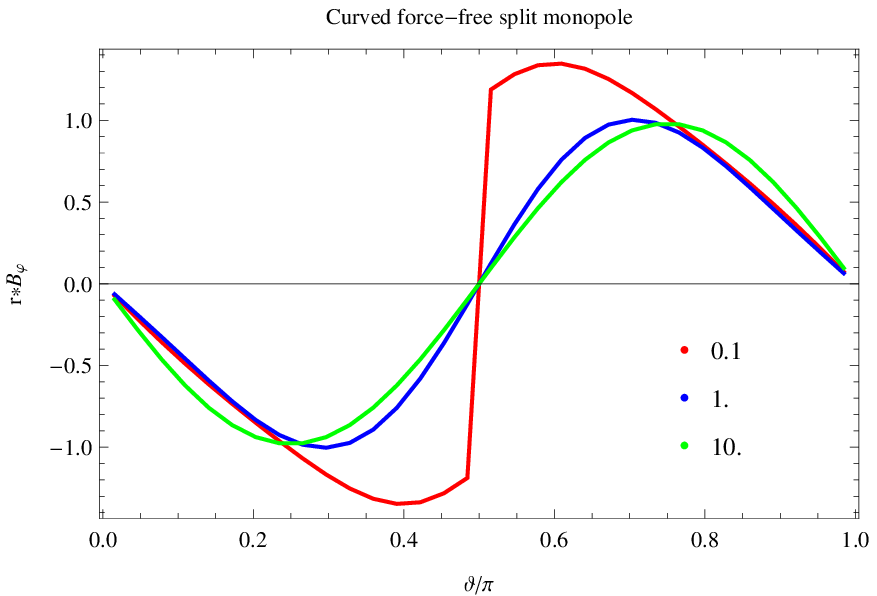} &
  \includegraphics[width=0.5\textwidth]{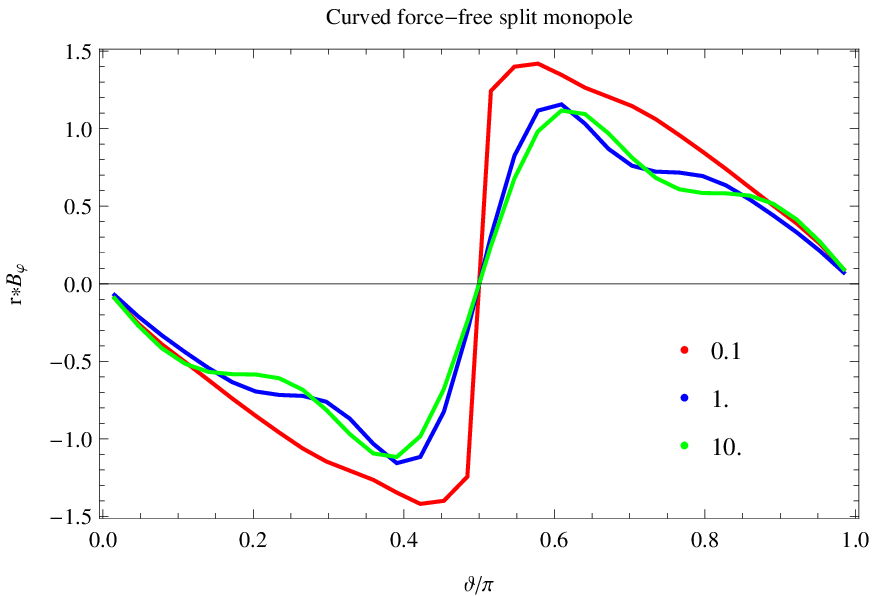}
  \end{tabular}
  \caption{Azimuthal component of the magnetic field~$B_\varphi$ for the split monopole solution, using different filtering orders, on the left, $\beta=4$ and on the right $\beta=8$. The parameters are $R_{\rm out}/\rlight = 10$ and $\rlight/R=10$. The radial location is the same as in fig.~\ref{fig:split_j1_B_p_flat}, $r=R$ for the red curve, $r=\rlight$ for the blue curve and $r=10\,\rlight$ for the green curve.}
  \label{fig:split_j1_B_p_curved}
\end{figure*}

We also show the azimuthal magnetic field component~$B_\varphi$ at three different radii, namely at the neutron star surface, at some point inside the simulation box and at the outer boundary,  fig.~\ref{fig:split_j1_B_p_curved}. The Gibbs phenomenon is apparent through its oscillations in the vicinity of the discontinuity. We gain accuracy by increasing the number of coefficients in the $\vartheta$ expansion and/or by increasing the order of the filtering. The discontinuity is better resolved by keeping higher orders but at the expense of introducing stronger oscillations. Compare the right panel ($\beta=8$) of fig.~\ref{fig:split_j1_B_p_curved} to its left panel ($\beta=4$).

\subsection{Performances}

As a last point, we show some performances of our code by providing the time needed for the code to produce the presented solutions for different resolutions and time steps. As an example, we looked at the time spend for computing the force-free monopole solution in curved space-time for the parameters $\rlight/R=2$ and $R_{\rm out}/\rlight=10$. Computations have been done on a single core processor with clock around 2.2~GHz. Results are summarized in table~\ref{tab:Temps}. For reasonable accuracy we need around one hour and for high precision a few days a needed. Note that for larger simulation boxes, such as the one presented in this paper, the computation time has to be multiplied by 10 for $R_{\rm out}=100\,\rlight$ or 100 for $R_{\rm out}=1000\,\rlight$. Another factor 5 is required if $\rlight/R=10$.

Such parameter space is at the edge of our current computational capability. We plane to write a new version employing the Message Passing Interface (MPI) in the near future to improve these performances and most importantly to be able to compute accurate three dimensional simulation of an oblique pulsar magnetosphere including the Fourier transform in the azimuthal direction.

\begin{table}
 \begin{center}
\begin{tabular}{llll}
\hline
Resolution & CFL=0.5 & CFL=0.2 & CFL=0.1 \\
\hline
\hline
$N_p \times K \times N_\vartheta$ & time (in s) & time (in s) & time (in s) \\
\hline
$1\times128\times 4$ & $4.6\times10^2$ & $1.2\times10^3$ & $2.3\times10^3$ \\
$1\times128\times 8$ & $8.9\times10^2$ & $2.2\times10^3$ & $4.4\times10^3$ \\
$1\times128\times16$ & $2.0\times10^3$ & $4.8\times10^3$ & $9.6\times10^3$ \\
$1\times128\times32$ & $4.9\times10^3$ & $1.2\times10^4$ & $2.4\times10^4$ \\
$1\times256\times 4$ & $1.9\times10^3$ & $4.7\times10^3$ & $9.6\times10^3$ \\
$1\times256\times 8$ & $3.5\times10^3$ & $8.8\times10^3$ & $1.8\times10^4$ \\
$1\times256\times16$ & $7.6\times10^3$ & $1.9\times10^4$ & $3.8\times10^4$ \\
$1\times256\times32$ & $1.8\times10^4$ & $4.6\times10^4$ & $9.1\times10^4$ \\
$1\times512\times 4$ & $7.5\times10^3$ & $1.9\times10^4$ & $3.8\times10^4$ \\
$1\times512\times 8$ & $1.4\times10^4$ & $3.4\times10^4$ & $6.9\times10^4$ \\
$1\times512\times16$ & $2.9\times10^4$ & $7.3\times10^4$ & $1.5\times10^5$ \\
$1\times512\times32$ & $7.1\times10^4$ & $1.8\times10^5$ & $3.6\times10^5$\\
\hline
 \end{tabular}
 \end{center}
\caption{Computational time (in seconds) for different resolutions and time steps (according to the Courant number~CFL) for the general-relativistic force free monopole field with $\rlight/R=2$ and $R_{\rm out}/\rlight=10$.}
\label{tab:Temps}
\end{table}

\section{CONCLUSION}
\label{sec:Conclusion}

General-relativistic force-free pulsar magnetospheres are the simplest approach to a self-consistent accurate investigation of the electromagnetic field configuration and plasma distribution around compact objects. In this paper, in order to quantify the effects of a curved background metric, we started with the force-free monopole field. We solved the three-dimensional time-dependent Maxwell equations in spherical geometry in the space-time of a slowly rotating neutron star. Approximate analytical monopole solutions in vacuum have been computed and successfully compared to the pseudo-spectral discontinuous Galerkin code. Then the force-free monopole field has been simulated. The corresponding spin-down luminosities remains very close to its flat space-time counterpart. We did not find any significant increase or decrease in the Poynting flux due to curvature and frame-dragging effects except for the high rotation rate given by $\rlight/R=2$ for which we found a decrease of several percent. The split monopole can also be computed in general relativity but the numerical stabilisation procedure remains too dissipative. This can be circumvent by increasing the spatial order of the method and the resolution of the grid.

Our next step will be to remove the monopole field assumption replacing it with the more realistic dipole field anchored in the neutron star. This allows a better quantitative description of the regions close to the neutron star surface. Knowing the plasma density and magnetic field structure at the polar caps is especially important for determining the location of the coherent radio emission. Phase-resolved radio polarisation and pulse profile emanating from those simulations will be very valuable observables to link with a wealth of radio astronomical data in the field. We hope that such study will help to constrain the inner magnetosphere of radio pulsar and sharpen our understanding of their low frequency emission properties.

\section*{Acknowledgements}

I am very grateful to the referee for his helpful comments and suggestions. This work has been supported by the French National Research Agency (ANR) through the grant No. ANR-13-JS05-0003-01 (project EMPERE). It also benefited from the computational facilities available at Equip@Meso of the Universit\'e de Strasbourg. I am grateful to Vasily Beskin for carefully reading the manuscript.


\label{lastpage}

\end{document}